\documentclass[preprint]{aastex}
\usepackage{aastexug}

\def\r{$r^\prime$}
\def\g{$g^\prime$}
\def\i{$i^\prime$}
\def\u{$u^\prime$}
\def\z{$z^\prime$}

\def\rr{$r^*$}

\def\kms{km s$^{-1}$}

\begin{document}

\title{Weak Lensing with SDSS Commissioning Data: The Galaxy-Mass Correlation
Function To $1h^{-1}$ Mpc}

\author{Philippe Fischer\altaffilmark{1}, 
Timothy A. McKay\altaffilmark{2},
Erin Sheldon\altaffilmark{2},
Andrew Connolly\altaffilmark{3},
Albert Stebbins\altaffilmark{4},
Joshua A. Frieman\altaffilmark{4,5},
Bhuvnesh Jain\altaffilmark{6},
Michael Joffre\altaffilmark{5},
David Johnston\altaffilmark{5},
Gary, Bernstein\altaffilmark{1},
James Annis\altaffilmark{4},
Neta A. Bahcall\altaffilmark{7},
J. Brinkmann\altaffilmark{9}, 
Michael A. Carr\altaffilmark{7}, 
Istv\'an Csabai\altaffilmark{6,10},
James E. Gunn\altaffilmark{7}, 
G. S. Hennessy\altaffilmark{11},
Robert B. Hindsley\altaffilmark{11},
Charles Hull\altaffilmark{12},
\v{Z}eljko Ivezi\'{c}\altaffilmark{7},
G. R. Knapp\altaffilmark{7},
Siriluk Limmongkol\altaffilmark{13},
Robert H. Lupton\altaffilmark{7}, 
Jeffrey A. Munn\altaffilmark{14},
Thomas Nash\altaffilmark{4},
Heidi Jo Newberg\altaffilmark{15},
Russell Owen\altaffilmark{13},
Jeffrey R. Pier\altaffilmark{14},
Constance M. Rockosi\altaffilmark{5}, 
Donald P. Schneider\altaffilmark{8}, 
J. Allyn Smith\altaffilmark{2},
Chris Stoughton\altaffilmark{4},
Alexander S. Szalay\altaffilmark{6},
Gyula P. Szokoly\altaffilmark{6},
Aniruddha R Thakar\altaffilmark{6},
Michael S. Vogeley\altaffilmark{16},
Patrick Waddell\altaffilmark{13},
David H. Weinberg\altaffilmark{17}, 
Donald G. York\altaffilmark{5}, (the SDSS Collaboration)
}


\altaffiltext{1}{University of Michigan, Department of Astronomy, 830 Dennison
Building, Ann Arbor, MI 48109}

\altaffiltext{2}{University of Michigan, Department of Physics, 500 East
University, Ann Arbor, MI 48109}

\altaffiltext{3}{Department of Physics and Astronomy, University of Pittsburgh,
Pittsburgh PA 15260}

\altaffiltext{4}{Fermi National Accelerator Laboratory, P.O. Box 500, Batavia,
IL 60510}

\altaffiltext{5}{University of Chicago, Astronomy \& Astrophysics
Center, 5640 S. Ellis Ave., Chicago, IL 60637}

\altaffiltext{6}{Department of Physics and Astronomy, The Johns Hopkins
University, 3701 San Martin Drive, Baltimore, MD 21218, USA}

\altaffiltext{7}{Princeton University Observatory, Princeton, NJ
08544}

\altaffiltext{8}{Department of Astronomy and Astrophysics, The Pennsylvania
State University, University Park, PA 16802}

\altaffiltext{9}{Apache Point Observatory, P.O. Box 59, Sunspot, NM 88349-0059}

\altaffiltext{10}{Department of Physics of Complex Systems, E\"otv\"os
University, P\'azm\'any P\'eter s\'et\'any 1/A, Budapest, H-1117, Hungary}

\altaffiltext{11}{U.S. Naval Observatory, 3450 Massachusetts Ave., NW,
Washington, DC 20392-5420}

\altaffiltext{12}{The Observatories of the Carnegie Institution of 
Washington, 813 Santa Barbara St, Pasadena, CA 91101}

\altaffiltext{13}{University of Washington, Department of Astronomy,
Box 351580, Seattle, WA 98195}

\altaffiltext{14}{U.S. Naval Observatory, Flagstaff Station, P.O. Box
1149, Flagstaff, AZ 86002-1149}

\altaffiltext{15}{Dept. of Physics, Applied Physics and Astronomy Rensselaer
Polytechnic Institute Troy, NY 12180}

\altaffiltext{16}{Department of Physics Drexel University
Philadelphia, PA 19104}

\altaffiltext{17}{Ohio State University, Dept.~of Astronomy, 174
W. 18th Ave., Columbus, OH 43210}

\slugcomment{Astronomical Journal, submitted}

\begin{abstract}

We present measurements of galaxy-galaxy weak lensing from 225 square degrees
of early commissioning imaging data from the Sloan Digital Sky Survey (SDSS).
We measure a mean tangential shear around a stacked sample of foreground
galaxies in three bandpasses ($g^\prime$, $r^\prime$, and $i^\prime$) out to
angular radii of $600^{\prime\prime}$, detecting the shear signal at very high
statistical significance.  The shear profile is well described by a power law
$\gamma_T = \gamma_{T0}(1\arcmin/\theta)^{\eta}$, with best fit slope of
$\eta=0.7-1.1$ (95\% confidence).  In the range
$\theta=10^{\prime\prime}-600^{\prime\prime}$, the mean tangential shear is
approximately $6 \pm 1 \times 10^{-4}$ in all three bands.  A variety of
rigorous tests demonstrate the reality of the gravitational lensing signal and
confirm the uncertainty estimates. In particular, we obtain shear measurements
consistent with zero when we rotate the background galaxies by $45^\circ$,
replace foreground galaxies with random points, or replace foreground galaxies
with bright stars.  We interpret our results by assuming that all matter
correlated with galaxies belongs to the galaxies. We model the mass
distributions of the foreground galaxies, which have a mean luminosity
$\langle$L$(\theta<5\arcsec)\rangle = 8.7 \pm 0.7 \times 10^{9} h^{-2}$
L$_{g^\prime\odot},\ 1.4 \pm 0.12 \times 10^{10} h^{-2}$ L$_{r^\prime\odot},\
1.8 \pm 0.14 \times 10^{10} h^{-2}$ L$_{i^\prime\odot}$, as approximately
isothermal spheres characterized by a velocity dispersion $\sigma_v$ and a
truncation radius $s$.  The velocity dispersion is constrained to be $\sigma_v=
150-190\;{\rm km}\;{\rm s}^{-1}$ at 95\% confidence ($145-195\;{\rm km}\;{\rm
s}^{-1}$ including systematic uncertainties), consistent with previous
determinations but with smaller error bars.  Our detection of shear at large
angular radii sets a 95\% confidence lower limit $s>140^{\prime\prime}$,
corresponding to a physical radius of $260h^{-1}$ kpc, implying that the dark
halos of typical luminous galaxies extend to very large radii. However, it is
likely that this is being systematically biased to large value by diffuse
matter in the halos of groups and clusters of galaxies. We also present a
preliminary determination of the galaxy-mass correlation function finding a
correlation length similar to the galaxy autocorrelation function and
consistency with a low matter density universe with modest bias.

The full SDSS will cover an area 44 times larger and provide spectroscopic
redshifts for the foreground galaxies, making it possible to greatly improve
the precision of these constraints, to measure additional parameters such as
halo shape and halo concentration, and to measure the properties of dark matter
halos separately for many different classes of galaxies.

\end{abstract}


\section{Introduction}

According to the theory of General Relativity, matter in the universe will
deflect rays of light. The consequence of this is that if one observes objects
located behind mass concentrations the background objects will appear
magnified, distorted and sometimes multiply imaged. This phenomenon, known as
gravitational lensing, can be used to measure the mass distributions of the
foreground objects. Mass measurements made in this manner are direct, and hence
free of the model-dependent biases of dynamical mass measurements.

Multiple imaging, strong distortions and large magnifications occur when the
light rays pass near very high mass density regions, and is referred to as
strong gravitational lensing. Regions which produce strong lensing are usually
near the centers of galaxies or clusters of galaxies. At larger impact
parameter, the distortions and magnifications are much smaller and multiple
images do not occur; this is known as the weak lensing regime. By studying the
small distortions in the background galaxies one can measure mass distributions
out to very large radii. Weak gravitational lensing has been successfully used
to measure mass distributions of clusters of galaxies (see the recent review by
\cite{me99}) and somewhat less successfully to measure halos of galaxies
\citep{ty84,br96,de96,gr96,hu98}. The problem in the latter case (known as
galaxy-galaxy lensing) is that an individual galaxy produces only a very small
distortion in the background galaxies. Since the background galaxies are
intrinsically elliptical the lensing signal is small compared to the ``shape
noise''.  Therefore, one needs to average the distortion behind many background
galaxies, which requires a large observed field.

The Sloan Digital Sky Survey (SDSS) will image 10000 square degrees in five
bandpasses. Although the images will be relatively shallow compared to many
previous weak lensing studies, they are, as we will show in this paper, well
suited to a galaxy-galaxy lensing study. Advantages of using brighter galaxies
include better known redshift distributions and larger angular size. The latter
means that the seeing requirements are less stringent than for studies based on
faint galaxies.

In this paper we report on a weak lensing study done with two nights of SDSS
commissioning data covering approximately 225 square degrees. We detect highly
significant weak lensing signals in the \g, \r\ and \i bandpasses. In \S
\ref{observations} we give details of the observations, in \S \ref{analysis} we
discuss our methods for analyzing the data including correcting for PSF
anisotropy (\S \ref{correct}), shear measurement (\S \ref{sheart}), and galaxy
redshift estimation (\S \ref{redshifts}). In \S \ref{error} we list sources of
random and systematic error and in \S \ref{tests} we discuss tests of the
significance of the shear measurements. In \S \ref{modeling} we model the data
and we present the discussion and conclusions in \S \ref{discussion} and \S
\ref{conclusions}. Throughout this paper we use $h={\rm H_0}/100\ {\rm km\ 
s^{-1}}$ and assume $\Omega = 1$. 

\section{Observations} \label{observations}

The SDSS 2.5m telescope is described by \cite{si00} 
(see also \\ {\tt
http://www.astro.princeton.edu/PBOOK/telescop/telescop.htm}). The telescope
provides a nearly 3$^\circ$, virtually undistorted field of view. The SDSS
imaging camera \citep{gu98,do00} is a mosaic of 30 $2048\times2048$ pixel CCDs
used for the primary imaging, along with an additional 24 500x2048 CCDs used
for astrometric and focus measurements. The imaging CCDs are arranged in six
columns of 5 CCDs each. Each of the 5 CCDs in a column views the sky through a
different broadband filter. The five filters (\u, \g, \r, \i, \z) \citep{fu96}
span a range from the atmospheric cutoff at 300nm to the limit of CCD
sensitivity at 1100nm.  Pixels in the imaging camera subtend
0.396$\prime\prime$ on the sky. Details of the data acquisition, reduction and
calibration can be found in \cite{pe00,uo00,ke00,pi00,tu00}.

SDSS imaging data are obtained in drift scan mode. In general, the telescope is
driven along a great circle on the sky in such a way that objects pass directly
down a column of CCDs. This allows essentially simultaneous observations to be
obtained in each of the five passbands and provides very efficient survey
observing (the shutter never closes).  Total integration time in each filter is
54.1s. Because the CCD columns are separated by nearly a CCD width, a single
SDSS observation of a strip of sky contains large gaps. The gaps in a single
``strip'' are then filled in on a subsequent night to obtain a completely
filled ``stripe''.  Successful drift scanning with a wide field system requires
an optical design with very low distortion, which has the added benefit of
removing an important shape systematic present for many other lensing studies.

The observations analyzed here consist of two nights of SDSS commissioning data
taken 20-21 March 1999 (SDSS runs 752 and 756). The total observation time was
seven hours the first night and eight hours the second night. These
observations were taken in a simplified mode in which the telescope remains
parked at the celestial equator. Atmospheric seeing averaged 1.5\arcsec\ (FWHM)
the first night and 1.25\arcsec\ the second night.  The RMS sky noise for these
data averages 26.0, 25.5 and 25.0 mag per square arcsec for \g, \r\ and \i,
respectively.

The portion of the data used in this analysis is a rectangle with corners (RA,
Dec.) = (9:40, -1:16) and (15:46, 1:16) (J2000) for a total of approximately
225 square degrees. Since the system sensitivity in \u\ and \z\ is relatively
low, only \g, \r\ and \i\ are used for determining the galaxy-galaxy lensing
signal.  The galaxy photometry was dereddened using the extinction values from
\cite{sc98}.

The photometric calibration used in this paper is only accurate to 5--10\%, due
to systematics in the shape of the point spread function across individual
CCDs, and the fact that the primary standard star network had not yet been
finalized at the time of these observations.  This situation will be improved
to the survey requirement of 2\% in the near future.  Thus we denote the
preliminary SDSS magnitudes presented here as $u^*$, $g^*$, $r^*$, $i^*$ and
$z^*$, rather than the notation $u'$, $g'$, $r'$, $i'$ and $z'$ that will be
used for the final SDSS photometry.




\section{Analysis} \label{analysis}

Weak lensing studies require accurate galaxy shape measurement for
gravitational shear estimation (see \S \ref{sheart}). The standard SDSS
pipeline software, outputs a plethora of parameters for each detected object
\citep{lu00} including accurate coordinates and photometry. For this paper, we
reprocess the data in order to measure galaxy shapes optimally as outlined in
\cite{be00}, which produces object (galaxy and star) shapes and
uncertainties. In order to do this, object coordinates and sizes are taken from
the pipeline output as starting guesses. We then measure quadratic moments
(Q$_{ij}$) of the surface brightness distributions weighted by an elliptical
Gaussian of location, size and orientation matched (through an iterative
procedure) to that of the object being measured:

\begin{equation}
Q_{ij} = \sum_{k,l}I_{k,l}G_{k,l}x_ix_j,
\end{equation}

\noindent where $I_{k,l}$ is the sky-subtracted surface brightness of pixel
$(k,l)$, $G_{k,l}$ is the value of the adaptively matched elliptical Gaussian
and $x_i$ is the pixel coordinate in the centroid-subtracted coordinate system.
From the quadratic moments one obtains the object ellipticities (e.g.,
\cite{ka95}):

\begin{eqnarray}
e_1 & = & {Q_{11} - Q_{22} \over Q_{11}+Q_{22}}, \\
e_2 & = & {2Q_{12} \over Q_{11}+Q_{22}}. \nonumber \\
\end{eqnarray}

\subsection{Correcting Galaxy Shapes} \label{correct}

The commissioning data used in this analysis were taken when the telescope was
not well collimated. Therefore, the image quality is relatively poor; the PSFs
are not radially symmetric and they vary strongly with both time and position
in the camera. This PSF anisotropy biases the shapes of galaxies. We attempt to
correct for anisotropic PSFs in the following manner. First we characterize the
PSFs in small regions spanning 600\arcsec $\times$ 800\arcsec\ by fitting
second order polynomials in R.A. and Dec. to the quadratic moments of stars.
We use stars with limiting magnitudes of 21, 20, and 19 (\g, \r\ and \i). We
then interpolate the polynomials at the positions of galaxies and correct the
measured galaxy shapes:

\begin{equation} \label{coreqn}
e_i(true)=e_i(measured)-S_{Sm}\times e_i(star),
\end{equation}

\noindent
where $S_{Sm}$ is the smear polarizability \citep{ka95,be00} which is
related to object size (and profile shape) relative to the local PSF.

\begin{figure}
\plotone{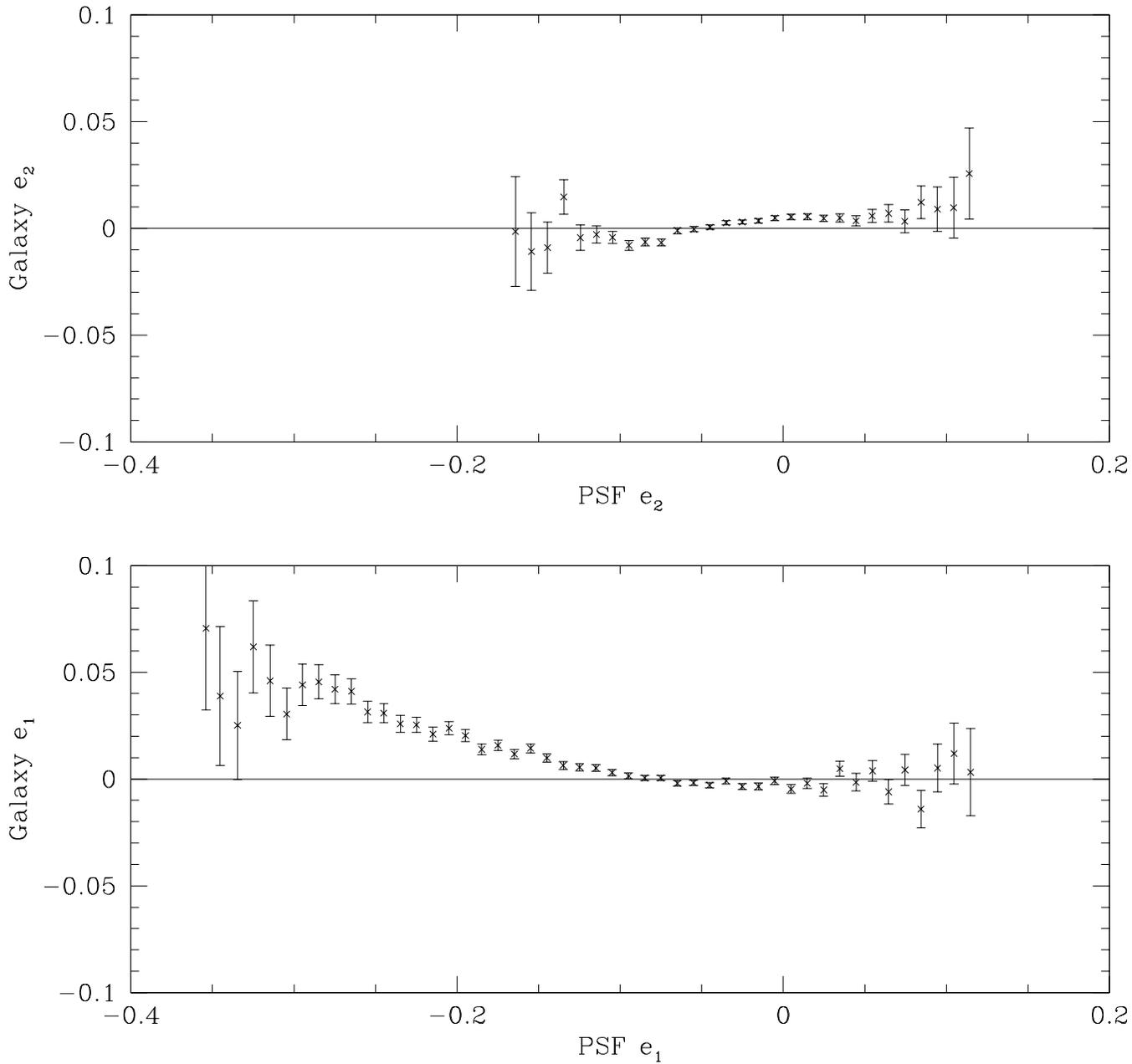}
\figcaption{Mean {\it corrected} galaxy shape vs. PSF shape for the \r\ data.
Strong correlations are present indicating imperfect correction for PSF
anisotropy.\label{galshapes}}
\end{figure}

Because the PSF shapes vary substantially during the observing period we can
test the efficacy of this method directly. Figure \ref{galshapes} shows a plot
of mean {\it corrected} galaxy $e_1$ and $e_2$ binned as a function of PSF
shape for all the data. One can see that after correction there are residual
systematic shape errors for the galaxies, especially where the initial PSF
anisotropy is large. This is probably because the correction in
Eqn. \ref{coreqn} is exact in the limit of nearly round PSF while much of our
data has PSF sufficiently elliptical to require higher order information about
the PSF.

Fortunately, for measuring a galaxy-galaxy lensing signal, image quality is
less important than for some other types of weak lensing measurements
(e.g. cosmic shear). The reason is that one measures tangential shear centered
at many different points in the field (see \S \ref{shearm}), hence, most
systematics will average out (see \S \ref{tests}). In order to minimize the
effects of systematically biased galaxy shapes we require the foreground
galaxies to be surrounded by a radially symmetric distribution of background
galaxies. For each foreground galaxy, the centroid of the background galaxy
distribution must not be significantly (4$\sigma$) different from the position
of the foreground galaxy and the distribution itself must have an ellipticity
of less than 20\% or deviate from circular by less than $3\sigma$ (whichever is
more). These requirements remove foreground galaxies which are near the edges
of the field or near regions that are lacking background galaxies (e.g.,
because of nearby bright stars or galaxies). By removing such galaxies we
reduce potential systematics as each galaxy will have its tangential shear
measured in a azimuthally symmetric manner.  In addition we fit linear
functions to the residual galaxy $e_1$ and $e_2$ shown in Figure
\ref{galshapes} and subtract the fit from the galaxy shapes.

The PSF will also blur the galaxies resulting in an underestimate of their true
ellipticities. This can be corrected by dividing the $e_i(galaxy)$ by
$1-S_{Sm}$. Objects with smear polarizabilities near unity (poorly resolved
objects similar in size to the PSF) will have a very large and uncertain
correction term. Thus we do not use any galaxies with $S_{Sm} > 0.8$ for our
shear measurements. This has the added benefit of eliminating the vast majority
of stars which could potentially contaminate our galaxy sample (see \S
\ref{systematic}).

When a galaxy image is subject to a shear its shape changes. The ratio of the
change in shape to twice the shear is known as the shear polarizability
\citep{ka95}. This is dependent on the shapes of the galaxies and needs to be
corrected for when measuring shear. We use the following formula for this
correction \citep{be00}:

\begin{equation}
S_{Sh}={\sum_i[w_i(1.-\sigma_{SN}^2 w_i e_T^2)] \over \sum_i{w_i}},
\end{equation}

\noindent where the $w_i$ are weights, $w_i = {1 \over
\sigma_{e,i}^2+\sigma_{SN}^2}$, $\sigma_{SN} = 0.32$ is the shape noise for
galaxies measured from a sample of large, high s/n SDSS images of galaxies,
$\sigma_e$ are seeing-corrected uncertainties \citep{be00}, and $e_T^2$ is the
tangential component of the ellipticity. We apply this correction statistically
to the final shear measurements (see equation \ref{gammaeqn}) (see \S
\ref{systematic} for details). We note that the galaxy shape uncertainties do
not include terms for centroiding error or background misestimation and are
therefore slightly underestimated.

\subsection{Shear - Theory} \label{sheart}

Weak gravitational lensing will cause galaxies located behind a mass
concentration to appear, on average, tangentially aligned with respect to the
center of mass. For gravitational lensing, the relationship between the
tangential shear, $\gamma_T$, and surface mass density, $\Sigma$, is
\citep{mi91,mi96},

\begin{equation}\label{escude}
\gamma_T(\theta) = \overline{\kappa}(\le \theta) - \overline{\kappa}(\theta),
\end{equation}

\noindent where $\kappa = \Sigma/\Sigma_{crit}$, the ratio of the surface
density to the critical surface density for multiple lensing, and $\theta$ is
the angular distance from a given point in the mass distribution. The critical
surface density depends on the redshift distribution of the background
galaxies. The first term on the right is the mean surface density interior to
$\theta$ and the second term is the mean surface density at $\theta$. In order
to measure the tangential shear around angular coordinates
$(\theta_1,\theta_2)$ the distortion of the background galaxies needs to be
measured:

\begin{equation}
D_i(\theta_1,\theta_2) = {1 \over
(1-S_{Sm,i})}{-(e_{1i}(\theta_{1,i}^2-\theta_{2,i}^2)+2e_{2i}\theta_{1,i}\theta_{2,i})
\over \theta_{1,i}^2+\theta_{2,i}^2},
\end{equation}

\noindent where $\theta_{1,i}$ and $\theta_{2,i}$ are angular distances in
R.A. and Dec., respectively, from the foreground galaxy to background galaxy
$i$. The mean tangential shear is:

\begin{equation} \label{gammaeqn}
\langle\gamma_T(\theta_1,\theta_2)\rangle = {1 \over
S_{Sh}}{\sum_i^{N_s}{w_iD_i} \over 2\sum{w_i}},
\end{equation}

\noindent where the weights, $w_i$ are described above and $N_s$ is the total
number of background galaxies in the region of interest. We do not use any
galaxies with seeing-corrected uncertainties of $\sigma_e > 0.64$. The
statistical uncertainty in the mean tangential shear, which mostly arises from
the intrinsic ellipticity and measurement error of the background galaxies (see
\S \ref{error}), is:

\begin{equation} \label{sigequation}
\sigma_{\gamma_T}^2(\theta_1,\theta_2) = {\sum_i^{N_s}{w_i^2D_i^2} \over
4S_{Sh}^2(\sum{w_i})^2},
\end{equation}

\noindent where we have assumed that $\langle D\rangle$ is zero, which, in the
weak lensing regime, results in a very minor overestimate in
$\sigma_{\gamma_T}$.

\subsection{Shear - Measurement} \label{shearm}

\begin{figure} 
\plotone{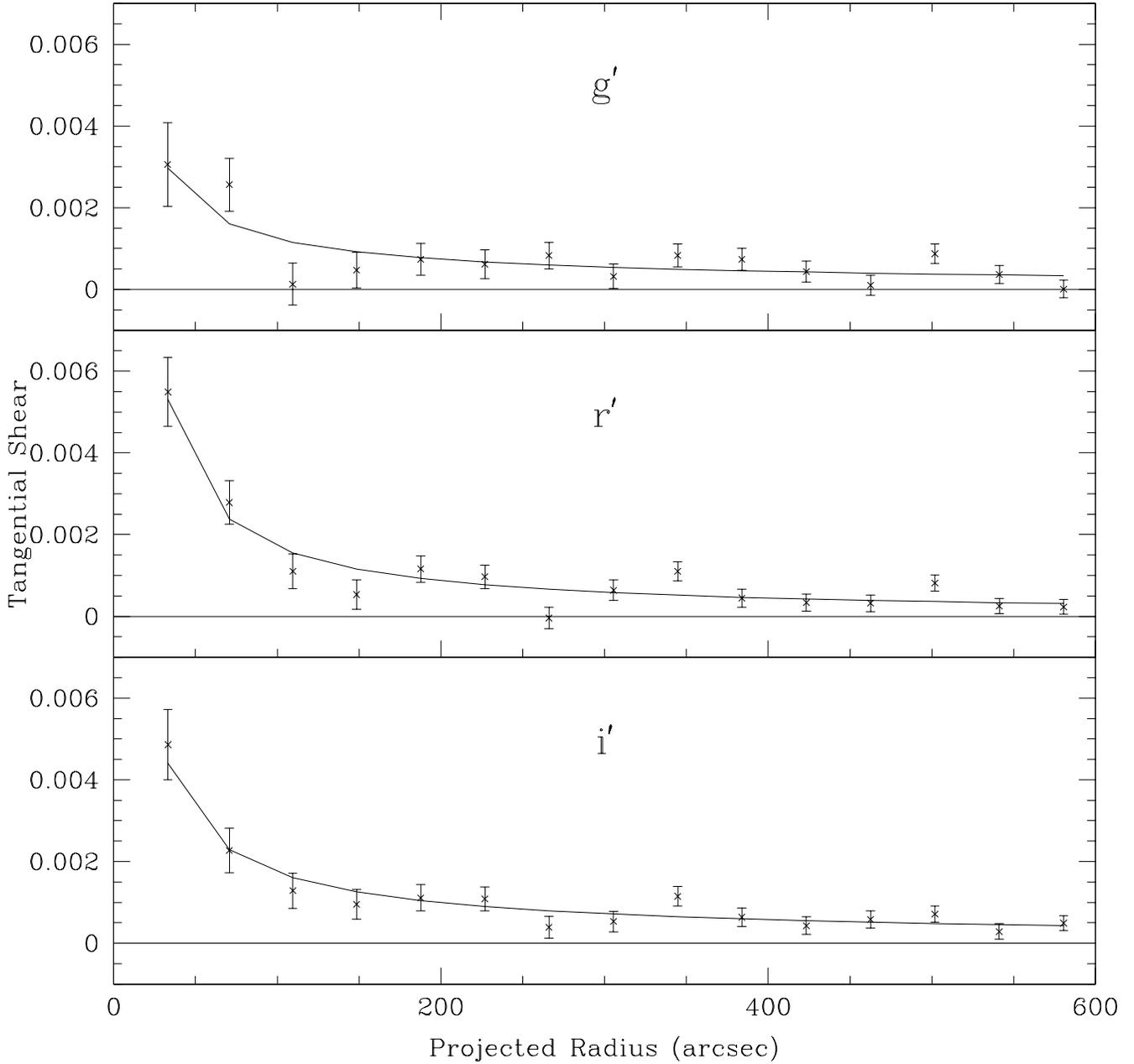} 
\figcaption{Mean shear around foreground galaxies measured from the g, \r and i
images (top, middle, bottom). The foreground galaxies have $16 \le $r$_0^* \le
18$ and the background galaxies have $18 \le $r$_0^* \le 22$. See Table
\protect\ref{tablea} for details. The solid lines are the best fit power-law
models with parameters given in Table \protect\ref{tableb}. The errorbars are
$\pm 1\sigma$ \label{shearr}}
\end{figure}

The shear from an individual L$_*$ galaxy is predicted to be small (less than
1\% for the present data). Therefore we need to average the shear around a
large number of foreground galaxies to obtain a statistically significant
signal.  Ideally one would like to select and scale the foreground galaxies
based on their redshifts. However, redshift information is not currently
available for the foreground galaxy sample (although the SDSS will eventually
measure redshifts for these galaxies) so we choose foreground and background
samples based on \r\ magnitudes.  Figure \ref{shearr} shows plots of measured
shear in the \g, \r, and \i\ images for $10\arcsec \le$ radius $\le
600\arcsec$. The lens or ``foreground'' galaxies have dereddened $16 \le
$\rr$_0 \le 18$ and the source or ``background'' galaxies have $18 \le $\rr$_0
\le 22$ (all magnitudes refer to Petrosian magnitudes, \cite{lu00}).  Columns
2, 3 and 4 of Table \ref{tablea} show the numbers of foreground galaxies,
background galaxies and foreground/background pairs, respectively, for each
filter. The \g and \i\ band have fewer galaxies because some of the \r-band
selected galaxies are not measurable in those bandpasses. Column 2 of Table
\ref{tableb} shows the mean values of the tangential shear for the radial range
10 - 600\arcsec.

\begin{figure} 
\plotone{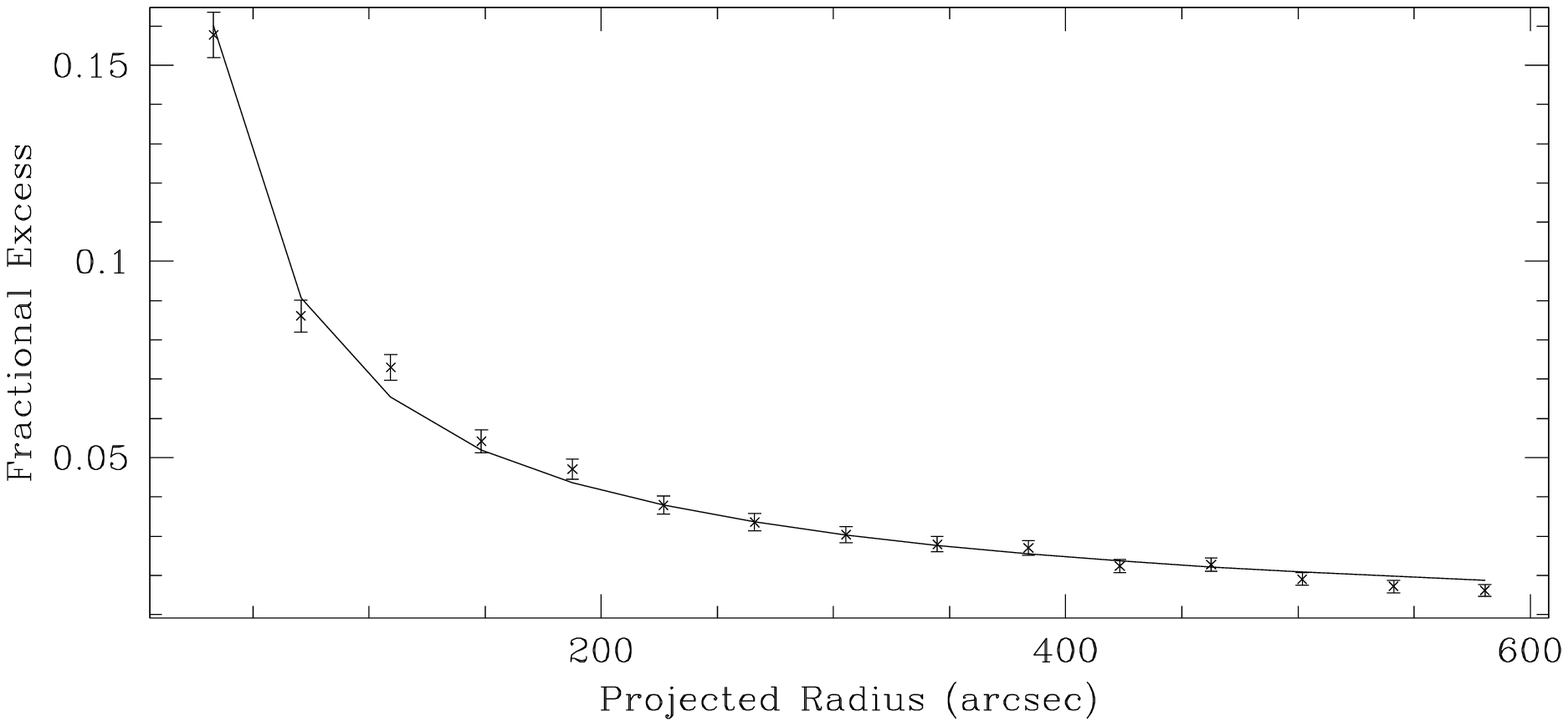} 
\figcaption{Density of ``background'' galaxies as a function of projected
radius from the ``foreground'' galaxies expressed as a fractional excess. The
solid line is a power law with index -0.75. The excess galaxies seen at small
radius are most likely associated with the lens galaxies. The diluting
influence of these galaxies must be corrected for in the shear
measurements. This plot is for \r\ data, \g\ and \i\ are similar. The errorbars
represent 1$\sigma$ Poisson errors.
\label{crosscor}}
\end{figure}

Our background galaxy sample contains many galaxies which are in front of some
of the foreground sample. This can be accounted for if the redshift
distributions of the foreground and background galaxies are known. However,
there is a complication due to the clustering of galaxies.  Some fraction of
the background galaxies will actually be galaxies associated with the lensing
galaxies and this fraction will decrease as a function of projected distance
from the foreground galaxies. If this correction is not made, the shear profile
will be radially biased.  Figure \ref{crosscor} shows the density of galaxies
in the background sample as a function of projected radius from the lens
galaxies for the \r\ data. The \g\ and \i\ data are similar. The shear values
in Figure \ref{shearr} have been corrected for this effect by multiplying the
measured values by the fractional excess of background galaxies. We assume that
there are no coherent distortions in the galaxies associated with the lensing
galaxies induced by dynamical interactions.

\subsection{Redshifts} \label{redshifts}

In order to convert the shear measurement into mass measurements we need to
know three things: 1) the foreground redshift distribution, $n(z_l)$, 2) the
background redshift distribution, $n(z_s)$ 3) and the weighting as a function
of redshift $w(z_s)$.  Although the SDSS will eventually measure redshifts for
the entire foreground galaxy sample, no spectroscopic redshifts are currently
available.  Therefore, we use photometric redshifts (photo-z) to estimate the
foreground and background galaxy redshift distributions.

The redshift distribution of the SDSS galaxy sample was determined from their
$u'$, $g'$, $r'$ and $i'$ photometric data (i.e.\ photometric-redshifts;
\cite{co95}). Redshifts are estimated using a semi-empirical template fitting
technique \citep{sa97,co99}. The quadratic difference between the observed
galaxy colors and those predicted by a set of model spectral energy
distributions are minimized as a function of redshift and galaxy type. This
minimization results in an estimated redshift, spectral type and associated
uncertainties. The spectral energy distributions used in this analysis are
based on the Coleman, Wu and Weedman data set, modified to match the observed
colors of the galaxies within the SDSS sample (see \cite{cs99,bu99} for a
discussion of these techniques).

To date there exist 1298 galaxies with published spectroscopic redshifts in the
equatorial region surveyed by Runs 752 and 756 (identified from the NASA/IPAC
Extragalactic Database and the spectroscopic survey of \cite{he97}). These
galaxies extend over the redshift and magnitude intervals of $0<z<0.5$ and
$r'<22$ respectively.  As such they provide an independent estimate of the
systematic and statistical uncertainties in the photometric redshift relation
for the SDSS commissioning data. A comparison between the spectroscopic and
photometric redshifts shows no systematic offsets in the photometric redshift
relation out to $z=0.5$ (the limit of the spectroscopic data). The dispersion
in the photometric redshift relation, at $r=21$, is $\sigma_z = 0.1$.

Photometric redshifts are currently available for only a portion of our data,
and in any case the photometric redshift errors are too large to compute
accurate values of $\Sigma_{crit}$ for each foreground-background galaxy
pair. Instead, we compute an expectation value of the quantity of interest
$\langle \Sigma^{-1}_{crit}\rangle$, assuming that these galaxies are drawn
randomly from the foreground and background redshift distributions, using an
$\Omega=1$ geometry to convert redshifts to physical distances.  Specifically,
we compute $\langle \Sigma^{-1}_{crit}\rangle$ from the following convolution:


\begin{equation}
\langle {\Sigma^{-1}_{crit}}\rangle = {1 \over
N_bN_f}\sum_{i=1}^{N_b}\sum_{j=1}^{N_f}{n_{b,i}n_{f,j} \over \Sigma_{crit,ij}},
\end{equation}

\noindent
where $N_b$ and $N_f$ are the number of foreground and background redshift
bins, respectively, $n_b$ and $n_f$ are the number of objects per redshift bin,
and $\Sigma_{crit,ij}$ is the critical density corresponding to $z_{f,j}$ and
$z_{b,i}$. The values of $\langle \Sigma^{-1}_{crit} \rangle$ for the three
filters are given in column 5 of Table \ref{tablea}. Also shown are the mean
values (weighted by $\langle \Sigma^{-1}_{crit} \rangle$) of the foreground
redshift and angular diameter distance.

\section{Sources of Error} \label{error}

\subsection{Random Error}

The errorbars plotted on the shear profiles in Figure \ref{shearr} are derived
via Equation (\ref{sigequation}) from the scatter in individual galaxy shape
measurements.  The two largest contributors to this scatter are the intrinsic
ellipticities of the background galaxies (shape noise) with an rms of 0.32, and
the error in the shape measurements ($e_1$, $e_2$) which is a function of
galaxy size and brightness. Improvements in image quality could reduce the
second source of scatter in future SDSS data, but the shape noise is an
intrinsic property of the galaxy distribution that is independent of image
quality.  The contributions of both these sources of scatter to uncertainties
in the shear decreases as $1/\sqrt{n_b}$, so the factor of $10,000/225=44$
increase in sky area in the full SDSS will greatly reduce the statistical
uncertainties in the shear measurements.



Another source of scatter arises because the foreground galaxies span a range
of redshifts and the shear measurements are binned by angle. Thus we are
averaging over a range of physical radii weighted by
$\Sigma^{-1}_{crit}(z_l)$. To estimate how much additional scatter this causes
we use the photo-z relationships from \S \ref{redshifts} to estimate n(z). We
assume that the galaxies are {\it identical} singular isothermal spheres and
find that at a given projected radius the standard deviation of the shear
arising from the lens redshift range is approximately 50\% of the mean
signal. Given that we have about 28000 foreground galaxies, this translates
into an uncertainty of 0.3\% and hence is a very small contributor to the
uncertainties in Figure \ref{shearr}.

A second source of scatter arises because the foreground galaxies do not have
identical masses (we are sampling a range of the galaxy mass function). In
order to estimate this we assign an absolute luminosity to each foreground
galaxy based on its apparent magnitude (reddening corrected) and photo-z. We
then assume that luminosity is proportional to mass squared. This produces a
scatter of about 40\% in mass which when averaged over all the galaxies
produces a similarly small contribution to the shear scatter.

A third source of scatter results from the possible radial asymmetry of the
galaxy surface density profiles. The shape of galaxy halos is not well known
(and is in fact something that can be measured with weak lensing, although it
will require about 10 times more data than we have here) but if we assume that
the halo has the same shape as the luminous portion of the galaxy (mean axis
ratio of about 0.5) then we can estimate the effect. To do this we assume a
pseudo-isothermal elliptical mass distribution \citep{ka93}. We then look at
tangential shear in two 90$^\circ$ angular bins, one centered on the major axis
and one centered on the minor axis. The ratio of mean shear in the minor axis
bin to the major axis bin is 0.55. Therefore, the scatter due to elliptical
halos is of similar order to the previous two effects.

We can, therefore, conclude that galaxy shape noise and measurement error are
the dominant sources of random uncertainty for our shear measurements.

\subsection{Systematic Error} \label{systematic}

We have already discussed how to correct for one source of systematic error,
the dilution of the shear signal due to including galaxies associated with the
lenses in the background sample. Additional systematics arise from: inclusion
of stars in the background, errors in the correction for blurring by the PSF
and errors in the estimate of $\Sigma_{crit}$.

Stars will dilute the shear signal by the stellar contamination
fraction. Distinguishing between stars and galaxies is difficult at the faint
end of our background sample as the two overlap in angular size. We eliminate
the vast majority of stars from our background sample by virtue of our smear
polarizability cut ($S_{Sm} \le 0.8$ for inclusion in the background
sample). We estimate the residual contamination by making a histogram of
$S_{sm}$ for all objects. This has two peaks, a narrow one centered at unity
due to stars and a broader one centered at 0.4-0.5 due to galaxies. We use this
to estimate that the residual contamination is less than 1\%.

Our estimate of $\Sigma_{crit}$ depends on knowledge of the redshift
distribution of the background and foreground galaxies. The error in
$\Sigma_{crit}$ depends on the relative error in the foreground/background
redshift distributions and the sensitivity of $\Sigma_{crit}$ to these errors.
We have estimated these distributions based on photometric redshifts.  The
leading source of error in the mean redshifts is uncertainty in the photometric
zeropoints for the commissioning data. Based on comparisons of the photometric
redshifts with approximately 500 spectroscopic redshifts the error in the mean
is about $\pm 0.04$. This will affect both foreground and background redshifts
in a similar manner and thus the effect on the value of $\Sigma_{crit}$ is
small; we conservatively adopt a 95\% confidence interval of $\pm 5\%$. 



To test our PSF blurring correction we carry out simulations of galaxy fields.
The simulated images have noise and seeing characteristics similar to that of
the \r-band SDSS images and the simulated galaxies have a similar distribution
of sizes.  Based on these simulations we conclude that our correction is
accurate to 10\%.

In conclusion we estimate the 95\% confidence limit on systematic errors to be
around $\pm 10\%$.

\subsection{Tests} \label{tests}

\begin{figure} 
\plotone{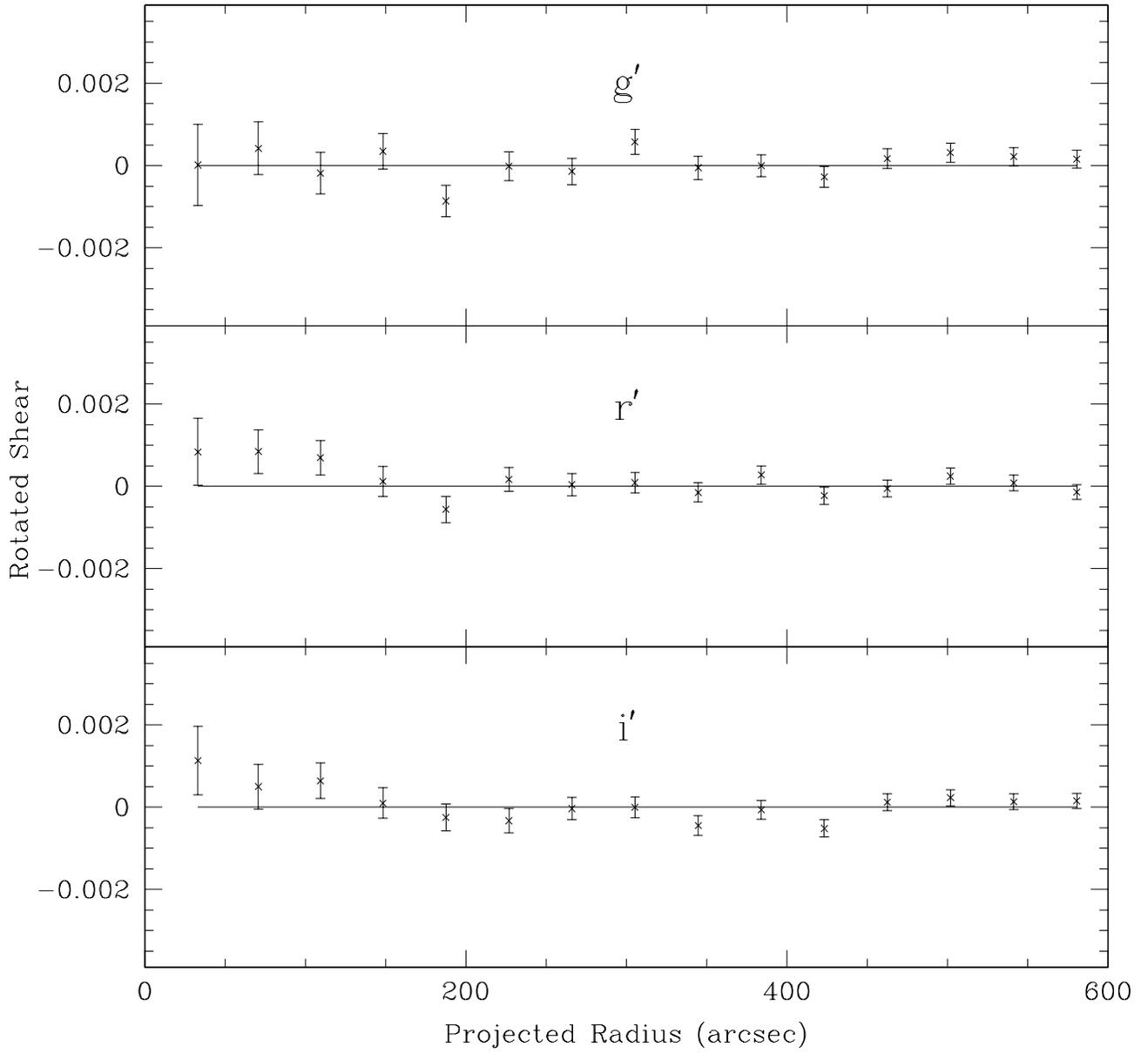} 
\figcaption{As in Fig. \protect\ref{shearr} but now the background galaxies
have been rotated by 45$^\circ$. The signal is consistent with zero for all
three bandpasses. The errorbars are $\pm 1\sigma$\label{radial}}
\end{figure}

In order to verify the reality of our shear detection we perform several
tests. The first test is designed to check if the signal is likely to be due to
gravitational lensing and involves rotating the background galaxies by
45$^\circ$ and determining if a signal still exists. This is equivalent to
calculating the curl of the gradient in $\kappa$ which should be zero for
lensing distortions \citep{st96,lu97}. Figure \ref{radial} shows the results of
this for the three bandpasses with the values given in column 3 of Table
\ref{tableb}. The mean ``rotated'' shear is consistent with zero for all three
cases.

\begin{figure} 
\plotone{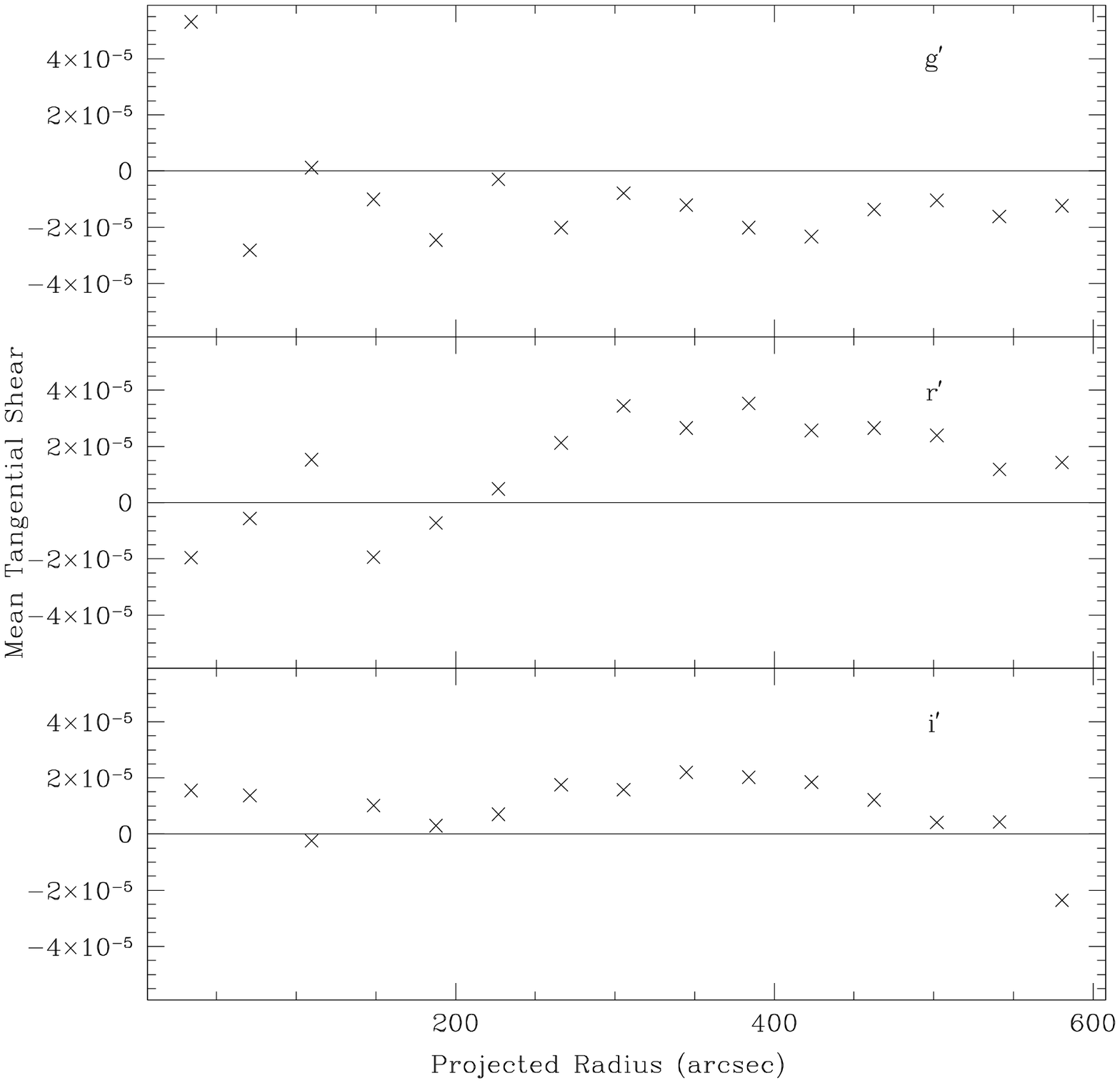} 
\figcaption{Mean shear around random points measured using the same background
galaxies as Figure \protect\ref{shearr}. The total random points are 1000 times
the number of true foreground galaxies. In the absence of systematics the
values should be zero. There is some evidence for systematics in each bandpass.
Note, the `Tangential Shear' axis has a range that is 1.5\% of the other shear
plots.
\label{random}}
\end{figure}

The second test is designed to check for systematics in the background galaxy
shapes and involves measuring shear around random points using the same
background galaxies as were used for the original shear measurement. Figure
\ref{random} shows a plot of the mean shear around 1000 sets of random points
for each bandpass. Each set contains as many points as the number of foreground
galaxies. The maximum absolute value for any of the radial points from these
simulations is less than 20\% of the corresponding errorbar in the observed
shear around the galaxies and the mean is 7\%. Even though we are using many
more random points than there are foreground galaxies the errorbars do not
continue to decrease as we are greatly oversampling the region. Thus it is not
possible to tell if this systematic is significant. Therefore, in the following
analysis, where appropriate, we quote results with and without the systematics
subtracted from our measured shear. In all cases, including the systematic does
not substantially change the result. 

\begin{figure} 
\plotone{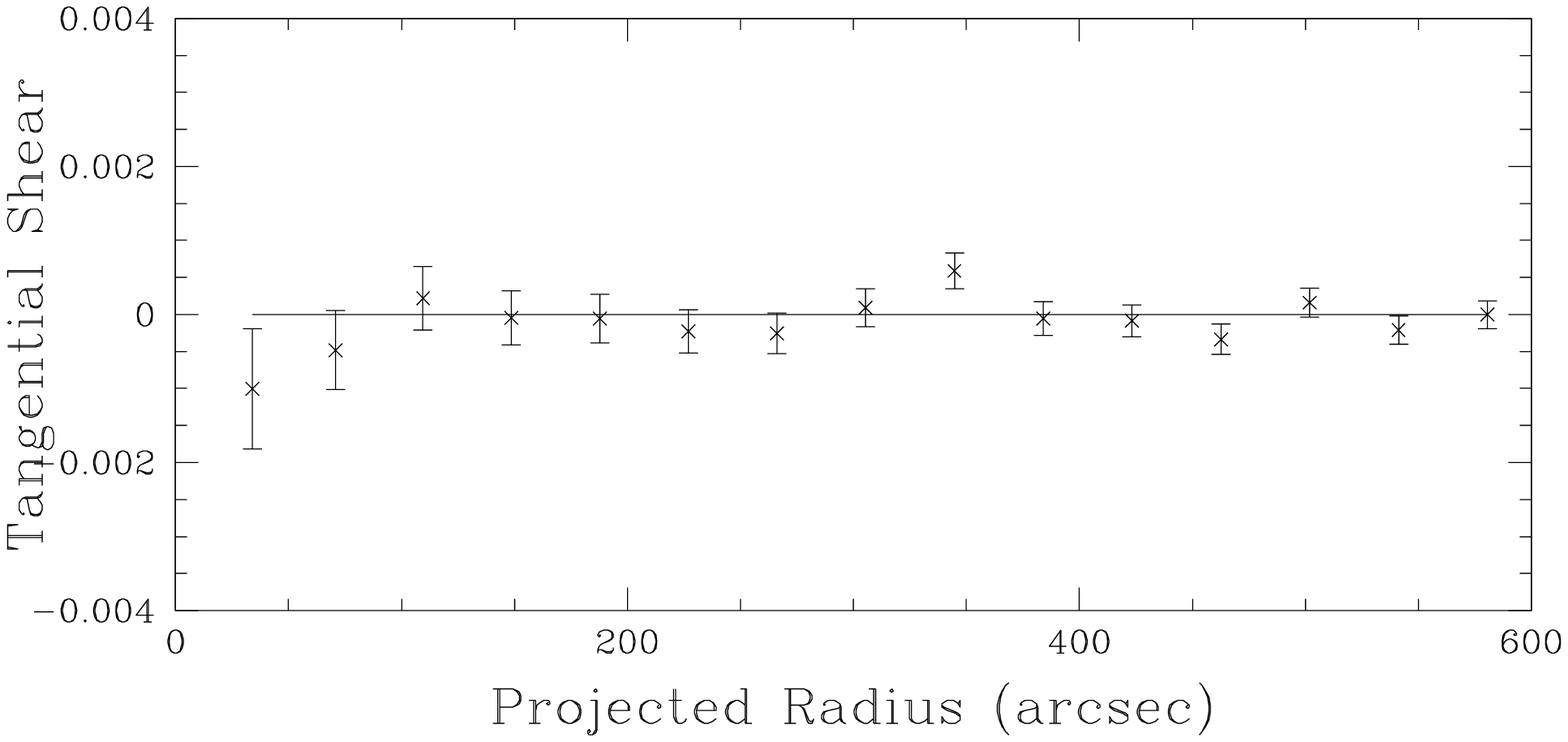} 
\figcaption{Mean shear around 26397 bright stars (\r-band). The signal is
consistent with zero implying that measurements biases due to gradients in the
surface brightness are not important. The errorbars are $\pm
1\sigma$\label{shearstars}}
\end{figure}

The third test is designed to check if the measured background galaxy shapes
are being biased by the faint tails of the foreground galaxies' surface
brightness distributions. For this test we measure shear around 26397 bright
stars (Figure \ref{shearstars}) and find $\langle \gamma_T \rangle = -3.6 \pm 5.5
\times 10^{-5}$, consistent with zero. Therefore, it is unlikely that gradients
in the sky background are resulting in significant systematic errors in the
galaxy shapes.

Based on the low level of possible systematic errors inferred from these three
tests we conclude that our shear measurement is real, results from
gravitational lensing and is not due to PSF systematics or measurement error.


\section{Modeling} \label{modeling}

\subsection{Galaxy-Shear Correlation Function}

\begin{figure} 
\plotone{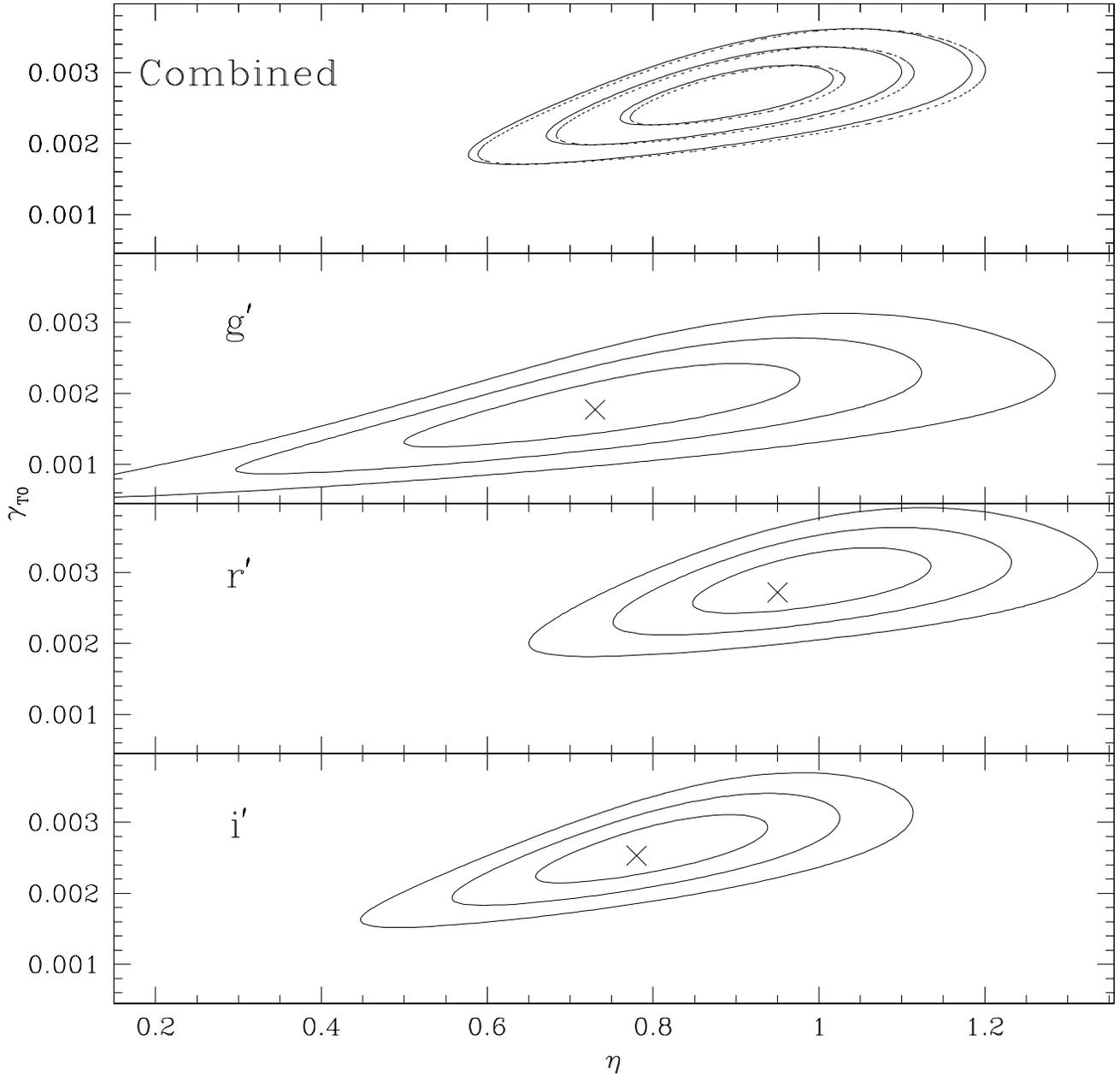} 
\figcaption{One, two and three $\sigma$ confidence intervals for the parameters
$\gamma_{T0}$ and $\eta$ for the \i, \r, and \g filters (bottom to second from
top) and the combined data (top) (see Equation \protect\ref{sheareqn}). The
dashed contours in the top panel show the confidence ranges if the systematic
errors shown in Figure \protect\ref{random} are subtracted from the shear
measurements prior to fitting. \label{shearconf}}
\end{figure}

The galaxy-galaxy shear signal is a direct measure of the galaxy-mass
correlation function (Figure \ref{shearr}). We fit power-law models of the form:

\begin{equation}\label{sheareqn}
\gamma_T(\theta)=\gamma_{T0} \left({1\arcmin\over\theta}\right)^{\eta},
\end{equation}


\noindent to each of the shear profiles where $\theta$ is the angular
radius. The best fit values are shown in Table \ref{tableb} and the confidence
intervals are shown in Figure \ref{shearconf}. The 95\% confidence range for
the slope from the three bandpasses combined (using the full error covariance
matrix) is $\eta = 0.7 - 1.1$. A shear profile well-fit by a power-law implies
the same power-law form for the surface mass density profile.

As the SDSS obtains more imaging data, we will be able to push this measurement
out to larger radii. It will then be interesting to compare this number to the
galaxy-galaxy correlation function (two-point correlation function) for a
direct measurement of the bias.

\subsection{Galaxy Parameters}

\begin{figure} 
\plotone{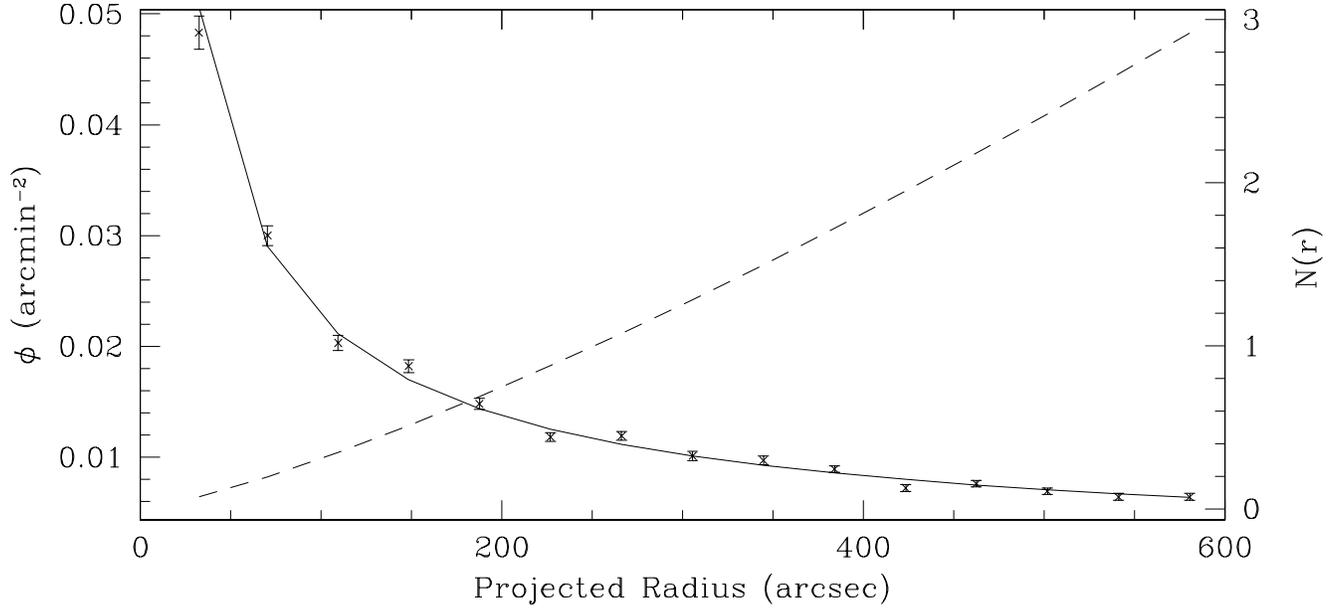} 
\figcaption{Density (points) and cumulative number (dashed line) of excess
``foreground'' galaxies as a function of projected radius from the central
galaxy. The solid line is a power-law with index -0.72. Both of these exclude
the central galaxy. These neighboring galaxies will contribute to the shear
signal.\label{galdens}}
\end{figure}

It is possible to constrain interesting galaxy parameters from these shear
measurements, however, this is complicated by the clustering of
galaxies. Because galaxies are correlated in space the measured shear profiles
have contributions from the central galaxies and neighboring galaxies.  As one
goes to larger projected radii, the fractional contribution of the neighboring
galaxies increases. To correctly infer the mean characteristics of the central
galaxies this correlation must be taken into account. Figure \ref{galdens}
shows the excess number density centered on foreground galaxies (foreground
galaxy autocorrelation function). The excess number density is well fit by a
function of the form:

\begin{equation}
\phi(\theta)=0.62 \left({1''\over\theta}\right)^{0.72} {\rm arcmin}^{-2}.
\end{equation}


Also shown in Figure \ref{galdens} is the cumulative number of neighboring
galaxies based on integrating the fit. Methods of dealing with the neighboring
galaxies have been discussed in \cite{br96,sc97,hu98}. These approaches involve
parameterizing the galaxy surface mass distributions and the use of scaling
relationships to account for variation in the lens galaxies. Maximum likelihood
is used to determine the most probable values of the galaxy parameters. These
approaches work best when there is at least crude redshift information
available. Lacking such information, we adopt a different and somewhat simpler
approach to modeling the shear.

The first contribution to the tangential shear is due to the central
galaxy. Similar to \cite{br96} and \cite{sc97} we approximate the mass density
of the lensing galaxies as a truncated isothermal:

\begin{equation}
\rho(R)={\sigma_v^2s^2 \over \pi GR^2(R^2+s^2)}
\end{equation}

\noindent where $\sigma_v$ is the line-of-sight velocity dispersion for $r<<s$,
and $s$ is the characteristic outer scale. For $R<<s$ this mass distribution
produces a near flat rotation curve as seen in observations of disk galaxies.
For $R>>s$ the profile falls as $1/R^4$ and the shear falls off very
rapidly. The corresponding surface density profile is:

\begin{equation} \label{surfdens}
\Sigma_g(\theta)={\sigma_v^2 \over 2G\theta}\left(1-{\theta \over
\sqrt{\theta^2+s^2}}\right),
\end{equation}



The second contribution is due to the neighboring galaxies. We assume that all
the galaxies have identical mass profiles given by equation (\ref{surfdens})
and that there is no contribution to the shear from galaxies not in our
foreground sample. The mass density profile resulting from the neighboring
galaxies is given by the convolution between the number density and the galaxy
surface density.

\begin{equation}
\Sigma_{tot}(\theta)=\left[\Sigma(\theta^\prime)*{\phi(\theta^\prime)}\right](\theta).
\end{equation}

\noindent and the shear is given by equation \ref{escude}.

\begin{figure} 
\plotone{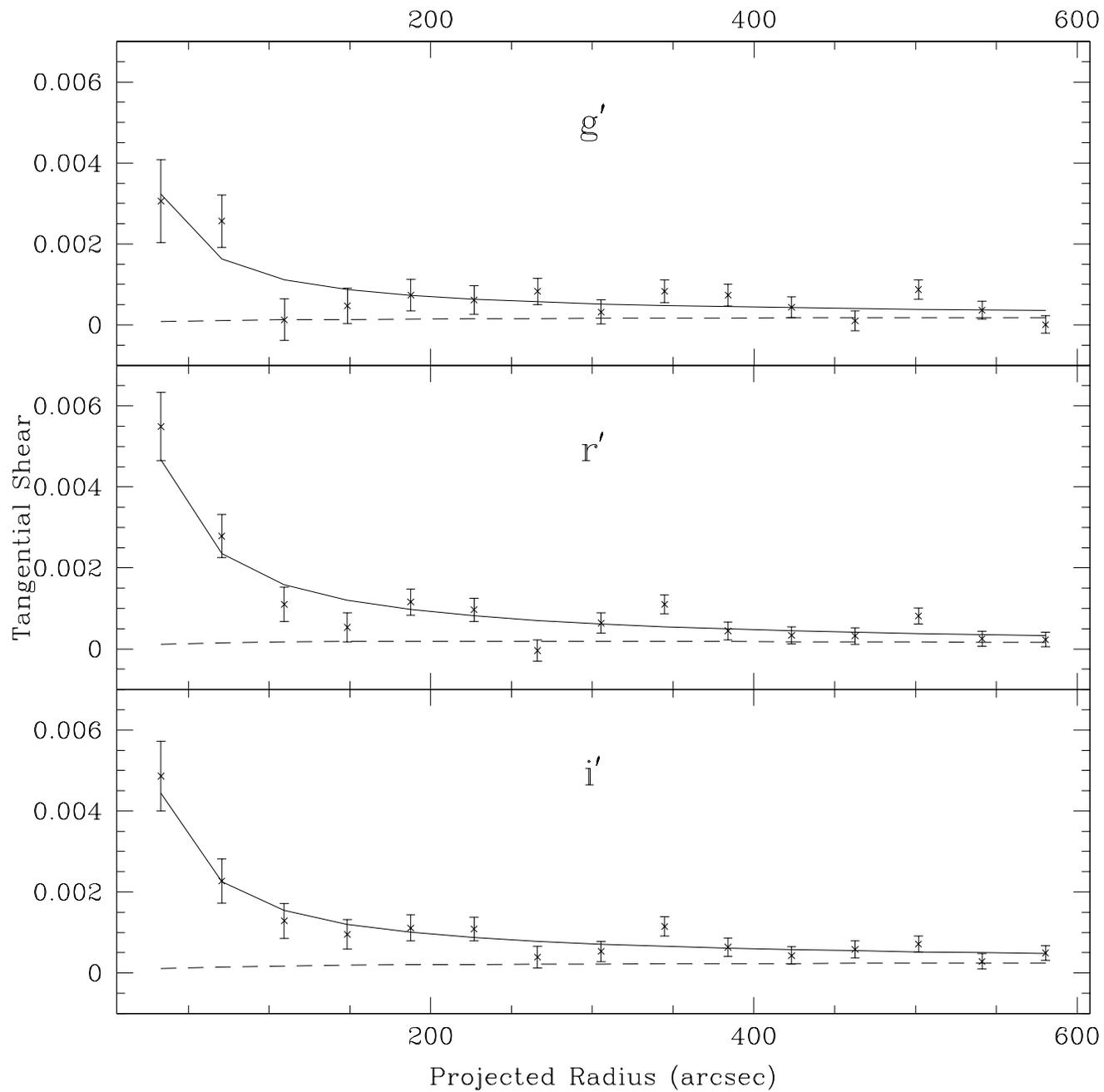} 
\figcaption{Shear around foreground galaxies as in
Figure \protect\ref{shearr}. The solid lines are the estimated shears due to
the central and neighboring galaxies. The dashed lines are the shears due to
the neighboring galaxies exclusively.\label{shearpred2}}
\end{figure}

\begin{figure} 
\plotone{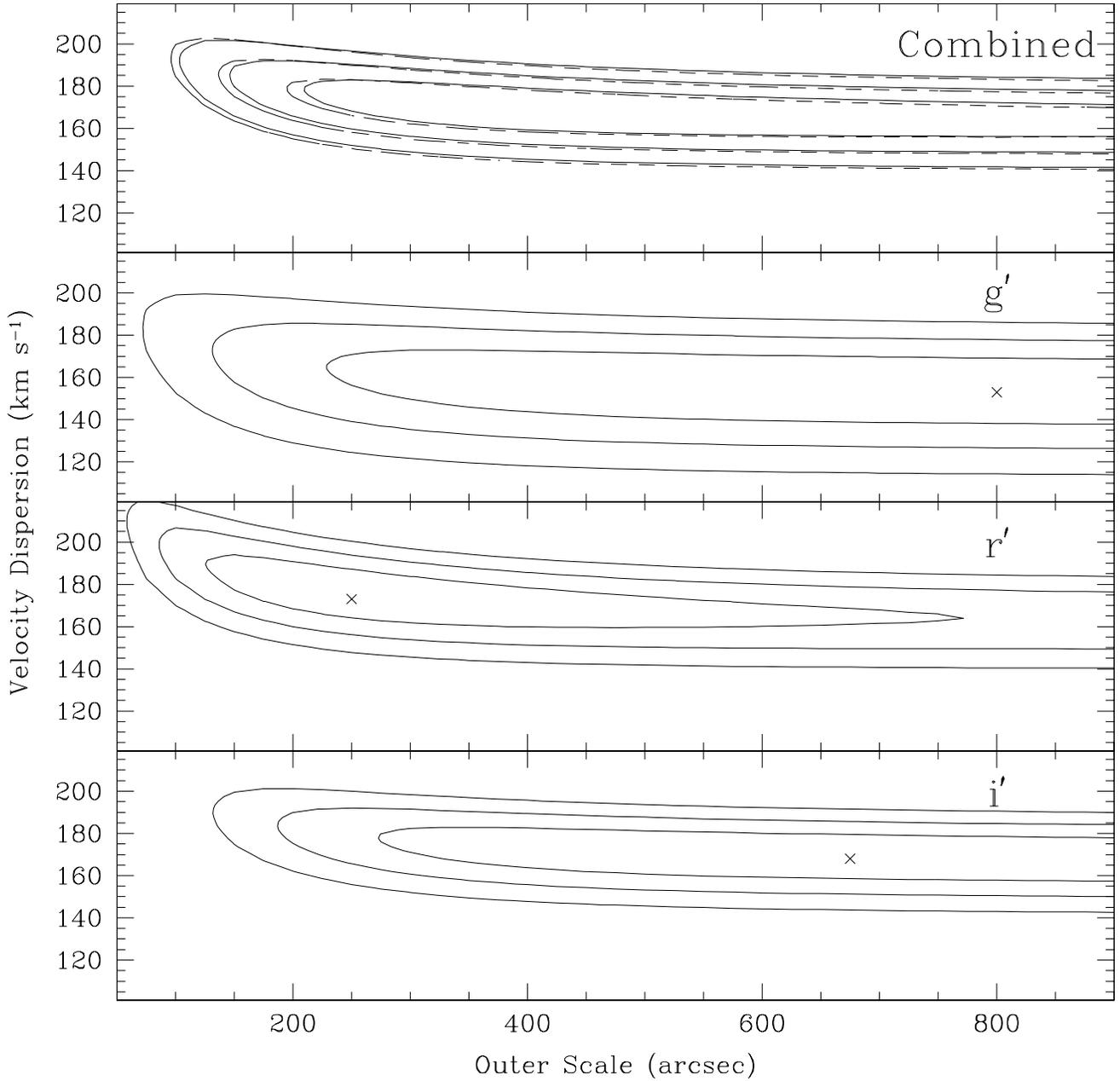} 
\figcaption{One, two and three $\sigma$ confidence regions for velocity
dispersion and $s$, the outer scale for \i, \r, and \g\ (bottom to second from
top) and for the combined data set (top). The points mark the best fit values.
The dashed contours in the top panel show the confidence ranges if the
systematic errors shown in Figure \protect\ref{random} are subtracted from the
shear measurements prior to fitting.
\label{modelconf}}
\end{figure}

We carry out this convolution numerically and use least-squares fitting to the
data shown in Figure \ref{shearr} to derive the best fit values for $\sigma_v$
and $s$ (columns 7 and 8 of Table \ref{tableb}). The values for $\sigma_v$
assume the $<D_l>$ shown in Table \ref{tablea}. The fits are shown in Figure
\ref{shearpred2}. In the inner regions the shear is completely dominated by the
central galaxy but the neighboring galaxies contribute roughly half the shear
at 600\arcsec. Figure \ref{modelconf} shows the 1, 2, and 3$\sigma$ confidence
regions for the fits to velocity dispersion and $s$ for the three bandpasses
and to the combined data set (using the full error covariance matrix). Also
shown are the confidence regions for the fits to the shears which have had the
systematics shown in Figure \ref{random} subtracted. The differences are very
small.

There are three important conclusions. The first is that the velocity
dispersion is well constrained (at least for $s \le 900\arcsec$) and is in the
range 150-190 km $s^{-1}$ (95\% confidence, 145-195 \kms\ if one includes
systematics). The second is that the minimum value for $s$ is well constrained
to be greater than 140\arcsec\ (95\% confidence) which is about 260 $h^{-1}$
kpc. The third conclusion is that the upper limit on $s$ is completely
unconstrained with the present data. The reason for this is that large changes
in $s$ mainly affect the large radii points and even substantial fractional
changes in the shear at large radii only change $\chi^2$ by a small amount.

One of the shortcomings of our analysis is that we do not account for the shear
contribution of neighboring galaxies fainter than our foreground sample
(Figure \ref{crosscor}).  If these galaxies contain a substantial fraction of
the mass then the consequence of ignoring these galaxies is to overestimate
both the outer scale radius and the velocity dispersion. Alternatively, one
could interpret these galaxies as actually being a constituent of the central
galaxies and then the inferred parameters ($\sigma_v$ and $s$) will include
their mass contribution.  We will consider their contribution in detail in a
future paper once we have spectroscopic and photometric redshifts and greater
field coverage.

\subsection{Galaxy-Mass Correlations}

Another way of looking at the tangential shear measurements is in terms of
2-point correlation functions.  The galaxy angular auto-correlation function,
$w_{\rm gg}(\theta)$, has been extensively studied, and the $\gamma_T(\theta)$
that we measure may be interpreted as the galaxy-shear correlation function,
which can be related to the spatial galaxy-mass correlation function, $\xi_{\rm
gm}(r)$ (see \cite{ka92}). Since the measured $\gamma_T(\theta)$ is well fit by
a power law in angle the inferred galaxy-mass correlation will also be well fit
by a power law in separation, i.e.

\begin{equation}
\xi_{\rm gm}(r)=\left({r_{\rm gm}\over r}\right)^\gamma
\end{equation}

\noindent
where $\gamma=\eta+1$, and $\eta$ is as given in Table \ref{tableb}.  For the
purpose of this analysis we assume a universal non-evolving correlation
function.  Of course different types of galaxies are known to have different
clustering properties and our foreground galaxy sample is not homogeneous,
containing more dim galaxies nearby and only luminous galaxies far away.  The
inferred value of $\gamma\approx 2$ is consistent with the clustering of
galaxies on the sub-Mpc scale we are probing, as is indicated by the slope of
the excess counts shown in Figure \ref{galdens}, which implies a galaxy
auto-correlation function

\begin{equation}
\xi_{\rm gg}(r)\approx\left({r_{\rm gg}\over r}\right)^{1.72} .
\end{equation}

\noindent
Thus a model where the galaxies trace the mass with a scale-independent bias is
consistent with our measurements.

To interpret the amplitude, and not just the slope, one must have knowledge of
the redshift distribution of the galaxies, both foreground and background.  As
we have mentioned this is somewhat uncertain, but we proceed in this section
using a model redshift distribution of background and foreground galaxies based
on the models of \cite{kgb93}. Using the Limber equation for $\gamma_T(\theta)$
(see \cite{ka92}) and imposing a power law model with slope $\gamma=1.72$,
taken from the excess counts, we infer for the correlation length of the
mass-galaxy cross-correlation

\begin{equation}
r_{\rm gm}\sim3h^{-1}\Omega_{\rm m,0}^{-0.57} {\rm Mpc}.
\end{equation}

\noindent
where $\Omega_{\rm m,0}$ is the density of clustering (mostly dark) matter in
units of the critical density.  Since we are mostly probing the mass-galaxy
correlations on sub-Mpc scales this inference involves an extrapolation of
nearly an order of magnitude in separation, and since the actual slope, $\eta$,
is not tightly constrained by our measurements this extrapolation is highly
uncertain. This uncertainty combined with the uncertainties in the actual
redshift distribution means that this value for $r_{\rm gm}$ should be
interpreted with great caution, and is mostly meant to put our results into a
familiar context. The fact that it is close to the correlation length of galaxy
clustering, $r_{gg}\approx5h^{-1}{\rm Mpc}$, is encouraging, although hardly
surprising since have obtained reasonable results for galaxy mass profiles.

Another approach is to compare $\gamma_T(\theta)$ with the galaxy clustering
implied by the excess counts of Figure \ref{galdens} (see \cite{vW98} for a
similar type of analysis).  The measurement error for the counts is much
smaller than for the shear, so we again assume the slope, $\gamma=1.72$,
implied by the excess counts.  We relate the mass-galaxy cross-correlation
function to the galaxy auto-correlation function by

\begin{equation}
\xi_{\rm gm}(r)={R\over b}\xi_{\rm gg}(r)
\end{equation}

\noindent
where $R$ and $b$ are assumed independent of separation. The bias factor, $b$,
is traditionally defined by the ratio of the rms inhomogeneities in galaxies
and in mass i.e. for scale-dependent bias $b=\sqrt{\xi_{\rm gg}(r)/\xi_{\rm
mm}(r)}$; but here we measure the cross-correlation and this may require an
additional parameter, $R$ (\cite{dl99,p98}).  By definition $R\in[-1,1]$, and
for linear bias $R=1$. Without going into details the measured ratio of
$\gamma_T(\theta)$ to the excess counts tells us that

\begin{equation}
\Omega_{\rm m,0}{R\over b}\approx0.3
\end{equation}

\noindent 
on the sub-Mpc scale.  The fact that $\Omega_{\rm m,0}$ with exponent 1
(approximately) appears in this equation rather than with a smaller exponent is
a consequence of the very small redshifts of the galaxies we are using.
Although this result does not require the large extrapolation needed to infer
$r_{\rm gm}$, we are still making assumptions about the redshift distributions
and the slope, so this result is only meant to be suggestive.  A more
quantitative analysis with less restrictive assumptions and going to larger
physical scales will follow, using more data, better redshift determinations,
and a proper analysis of $w_{\rm gg}(\theta)$.  Our preliminary result is
consistent with currently fashionable cosmological models with low $\Omega_{\rm
m,0}$ and moderate $b$, but using our measurements we cannot disentangle the
relative contribution of the three parameters, $\Omega_{\rm m,0}$, $b$ and
$R$. In the future lensing studies of SDSS data may be able to measure the
shear-shear correlations at low redshifts which would allow us to determine
$\Omega_{\rm m,0}/b^2$, and studies of the strength of lensing with redshift
may be able to disentangle $\Omega_{\rm m,0}$ from $R/b$, just using lensing
from SDSS.

\section{Discussion} \label{discussion}

Our value for the velocity dispersion is consistent with the galaxy-galaxy
lensing measurements of \cite{br96} ($\sigma_v= 155 \pm 55$ \kms), \cite{de96}
($\sigma_v = 185^{+30}_{-35}$ \kms), and \cite{hu98} ($\sigma_v = 150 \pm 30$
\kms) (all uncertainties are $\pm 1\sigma$). All three of these studies used
lens galaxies at much higher redshifts than considered here (particularly the
last two which utilize the Hubble Deep Field \citep{wi96} and had mean lens
redshift of around 0.6) and were measuring shear at much smaller radii than
considered here.

\begin{figure} 
\plotone{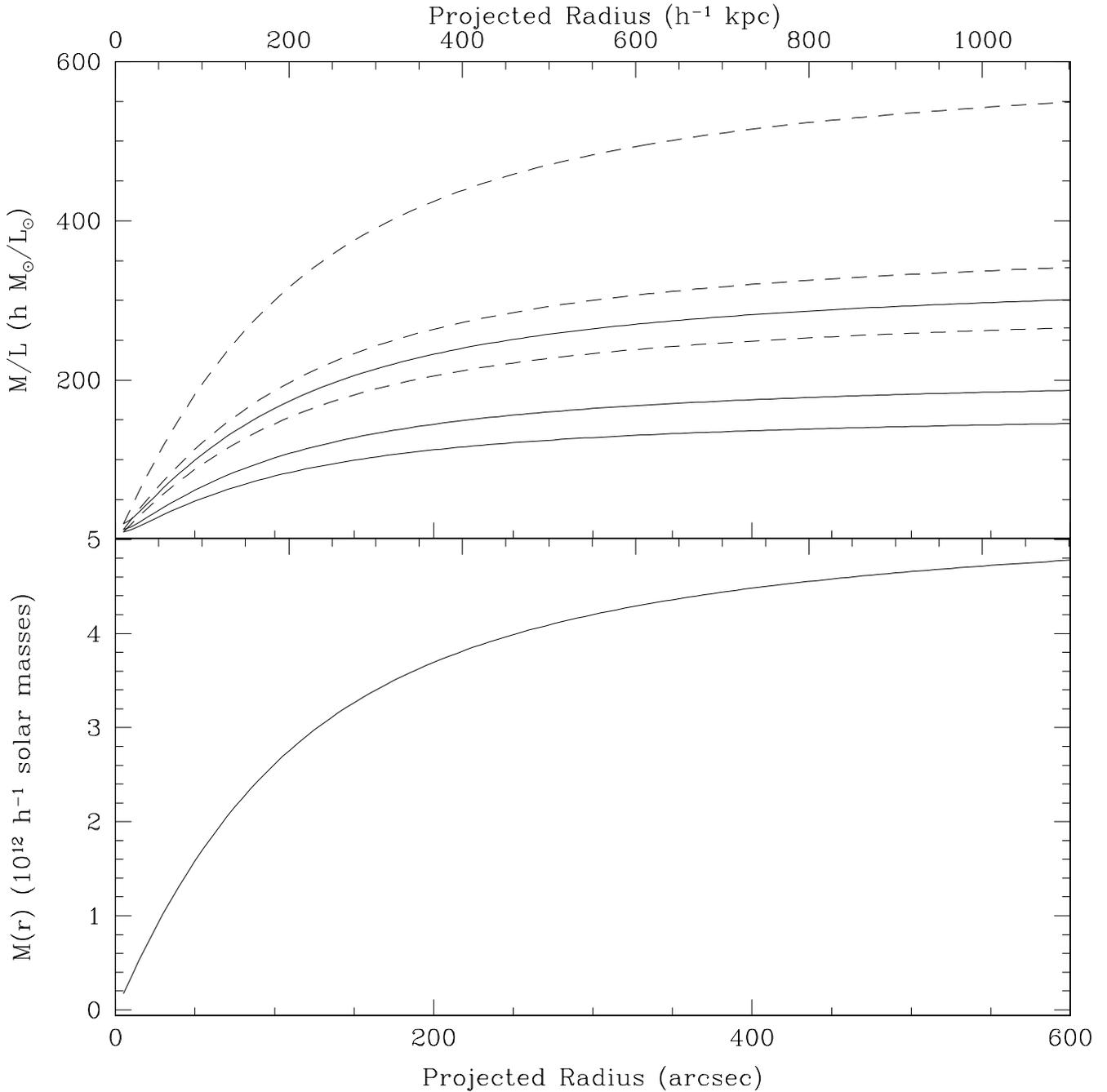} 
\figcaption{Cumulative mass (lower) and mass-to-light profiles (upper) for
$\sigma_v = 170$ \kms\ and $s = 140\arcsec$.  The dashed M/L profiles assume
all the galaxy light is within 5\arcsec\ of the center (as measured) while the
solid lines extrapolate the light profile as described in the text. The
profiles are \g, \r, \i\ from top to bottom. This represents the minimal mass
model (95\% lower limit on $s$) so the true profiles are likely to have higher
values.\label{mass}}
\end{figure}

The 95\% confidence lower limit on the outer scale radius (260 $h^{-1}$ kpc)
implies that there is a great deal of mass at large radius in galaxies.  In
Figure \ref{mass} we plot a cumulative mass distribution for $\sigma_v = 170$
\kms and $s = 140\arcsec$. Even for this minimum mass case we find
M(r)$\approx5\times 10^{12} h^{-1}$ M$_\odot$ for $r=600\arcsec$.  The mean
mass profile is averaged over galaxies spanning a wide range of environments,
varying from low density regions to dense cluster centers. Therefore, what
exactly do the inferred galaxy parameters mean? For example if, as indicated by
some recent observations, clusters have large mass components which are not
associated with individual galaxies \citep{ty98} this mass will be included in
our shear measurements. The galaxy-galaxy shear signal is a measure of all the
mass correlated with galaxy positions (even if the mass is not gravitationally
bound to the galaxies). The inferred mean profile is, therefore, based on a fit
to all the mass, both bound and unbound, which is correlated with the galaxy
positions and is a measure of the mean mass per galaxy over a representative
sample of the universe. We note that if there is a large mass component in the
universe which is not correlated with galaxy positions then our procedure
underestimates the mass per galaxy. \cite{za94} measured the mass within $150
h^{-1}$ kpc for a sample of isolated disk galaxies with rotation velocities
near 250\kms\ obtaining M = $1.1 - 2.0 h^{-1} \times 10^{12}$ M$_\odot$ (90\%
confidence). The upper limit is similar to the value plotted in Figure
\ref{mass} although direct comparison is dubious given the different ways the
galaxies were selected. Both results, however, imply a large extent to the
halo.

An important value is the ratio of galaxy mass to light (M/L) as one can use it
to estimate the density parameter $\Omega_{galaxies}$. The SDSS photometric
pipeline yields galaxy photometry within the Petrosian radius \citep{lu00}
which have a mean value of 5\arcsec\ for the data used here. Using the photo-z
measurements described in \S \ref{redshifts} we find
$\langle$L$(\theta<5\arcsec)\rangle = 8.7 \pm 0.7 \times 10^{9} h^{-2}$
L$_{g^\prime\odot},\ 1.4 \pm 0.12 \times 10^{10} h^{-2}$ L$_{r^\prime\odot},\
1.8 \pm 0.14 \times 10^{10} h^{-2}$ L$_{i^\prime\odot}$. Figure \ref{mass}
shows the cumulative mass-to-light profile for the minimal galaxy
($s=140\arcsec$) assuming all the galaxy light is contained within
5\arcsec. Also shown is the cumulative M/L profile where we have extrapolated
the light profile by assuming it is an exponential with scale length similar to
that of the Milky Way (5 kpc \cite{gi90}). Values of M$_{\odot} = 5.06, 4.64,
4.53$ \citep{fu99} for \g, \r, and \i, respectively, have been used.  The
region within $100h^{-1}$ kpc has M/L$_r^\prime \approx 50 h$
M$_\odot$/L$_{r^\prime,\odot}$ and for the largest radii measured ($\sim 1
h^{-1}$ Mpc) it rises to a value similar to what is seen in clusters. These
results are similar with the observed dependence of M/L on scale obtained by
\cite{ba95} using observations of galaxies, groups and clusters.  For higher
values of $s$ the M/L is correspondingly higher, particularly at large
radii. Note that we have not made k or evolution corrections and include only
light associated with galaxies (e.g., we do not include possible contributions
from intracluster light).


\section{Conclusions and Future Work} \label{conclusions}

In this paper we have presented the first weak lensing results from the Sloan
Digital Sky Survey. One of our most important conclusions is that weak lensing
studies are possible with the SDSS despite the fact that it is a relatively
shallow drift-scan survey. The data analyzed here are 225 square degrees of
early commissioning images which suffered from poor image quality. Despite this
we detect a galaxy-galaxy lensing signal around a large sample of foreground
galaxies in three bandpasses (\g, \r\ and \i) at very high significance.

We present power-law fits to the shear signal and find slope values ranging
from 0.7-1.1 (95\% confidence range). Our attempts to determine galaxy
parameters yielded a galaxy velocity dispersion of $\sigma_v = 150 - 190$ \kms
(95\% - 145 - 195 \kms including systematics) and a 95\% lower limit of
140\arcsec\ (260 $h^{-1}$ kpc) for the outer scale radius. It should be noted
that these parameters may include significant contributions from inter-galactic
mass provided this mass is correlated with galaxy positions. This lower limit
on $s$ leads to a value of M/L$_r^\prime \ge 200 h$
M$_\odot$/L$_{r^\prime,\odot}$ within a projected radius of 1 $h^{-1}$ Mpc.

While our results can be used to constrain the profile and velocity dispersion
of galactic halos, at large scales it is more meaningful to use them as a
measure of the projected galaxy-mass correlation function. The dividing line
between galactic halos and large-scale structure is somewhat arbitrary, but a
rough guide is provided by the fact that galactic halos cannot exist as
virialized structures on scales much larger than 100 $h^{-1}$ kpc, since the
dynamical time becomes larger than the age of the universe. We present a
preliminary determination of the galaxy-mass correlation function finding a
correlation length similar to the galaxy autocorrelation function and
consistency with a low matter density universe with modest bias. However,
degeneracies between $\Omega$ and bias complicate the interpretation.

It will be useful to compare our future results on larger scales with
theoretical models of the galaxy-mass correlation.  The lensing strength
depends on the amplitude of fluctuations in the absolute mass density
associated with galaxies. Therefore by comparing with theoretical predictions,
or through a direct inversion of the data, we will be able to constrain the
biasing of galaxies, the mean density parameter $\Omega$, and the dark matter
power spectrum. The accuracy with which these parameters can be independently
constrained will be aided by the increased sample size, redshift information,
and combination with other measures such as the angular clustering of galaxies
and probes of field lensing. We will explore these approaches in future work by
using simulated mock catalogs as well as analytical models of halo-mass
correlations \citep{mc77,sh97}.

Ongoing improvements in the SDSS data will greatly enhance galaxy-galaxy
lensing studies. First, the image quality should improve as the telescope
optics are designed to yield PSFs with ellipticities of less than 12\%. Second,
we will have detailed spectra for all foreground galaxies and photometric
redshifts for a large fraction of our background sample. This will allow us to
put all the foreground galaxies on the same physical scale and will greatly
improve our ability to interpret the galaxy-galaxy lensing measurement in terms
of galaxy parameters. The third is that the coverage will increase by a factor
of fifty over what we have presented here, which, if our uncertainties remain
dominated by random errors, implies a signal-to-noise improvement by a factor
of seven. 


Finally, if the SDSS achieves its imaging design goals, other types of weak
lensing studies will also be possible \citep{go94,st96}. These include the
search for mass overdensities and measurement of the cosmic shear. Of
particular interest is the 250 square degree Southern survey, where repeat
scans will produce images two magnitudes fainter than the main survey.

\acknowledgments 

The Sloan Digital Sky Survey (SDSS) is a joint project of The University of
Chicago, Fermilab, the Institute for Advanced Study, the Japan Participation
Group, The Johns Hopkins University, the Max-Planck-Institute for Astronomy,
Princeton University, the United States Naval Observatory, and the University
of Washington. Apache Point Observatory, site of the SDSS, is operated by the
Astrophysical Research Consortium. Funding for the project has been provided by
the Alfred P. Sloan Foundation, the SDSS member institutions, the National
Aeronautics and Space Administration, the National Science Foundation, the
U.S. Department of Energy and the Ministry of Education of Japan. The SDSS Web
site is http://www.sdss.org/. Tim Mckay acknowledges support through NSF PECASE
AST9703282.

\pagebreak

\begin{deluxetable}{ccccccc}
\tablecolumns{7}
\tablewidth{0pt}
\tablecaption{Data}
\tablehead{
 \colhead{Filter} & \colhead{Foreground} & \colhead{Background} &
 \colhead{Pairs} & \colhead{$<\Sigma^{-1}_{crit}>$} & \colhead{$<z_l>$} &
 \colhead{$<D_l>H_0/c$} \\
  & & & & g$^{-1}$ cm$^2$ & }
\startdata
  \g & 28134 & 1213862 & 13471567 & 0.343 & 0.168 & 0.123 \\
  \r & 27890 & 1552721 & 16913287 & 0.392 & 0.172 & 0.126 \\
  \i & 27945 & 1447336 & 15809470 & 0.403 & 0.173 & 0.126 \\
\enddata 
\label{tablea}
\end{deluxetable}

\begin{deluxetable}{ccccccccc}
\tablewidth{0pt} \tablecaption{Model Fits} \tablehead{ \colhead{Filter} &
\colhead{$\langle\gamma_T(10-600\arcsec)\rangle$} &
\colhead{$\langle\gamma_R(10-600\arcsec)\rangle$} & \colhead{$\eta$\tablenotemark{a}} &
\colhead{$\gamma_{T0}$\tablenotemark{a}} & \colhead{P($>\chi^2$)} & \colhead{$\sigma_v$\tablenotemark{a}} &
\colhead{$s$\tablenotemark{a}} & \colhead{P($>\chi^2$)} \\
& ($\times 10^{-4}$) &($\times 10^{-5}$) & & ($\times 10^{-5}$) & & (km s$^{-1}$) & (arcsec) }
\startdata 
\g & $5.2 \pm 0.77$ & $~~7.3 \pm 7.5$ & 0.73(0.70) & 177(181) & 0.12 & 153(155) & 800(975) & 0.13 \\ 
\r & $5.8 \pm 0.64$ & $~~3.6 \pm 6.4$ & 0.95(0.97) & 272(268) & 0.03 & 173(181) & 250(200) & 0.02 \\ 
\i & $6.9 \pm 0.65$ & $-1.1\pm 6.5$   & 0.78(0.79) & 254(252) & 0.54 & 168(168) & 675(700) & 0.59 \\
\enddata
\tablenotetext{a}{Values in parenthesis refer to fits to the shears with
the systematics from Figure \ref{random} subtracted.}
\label{tableb}
\end{deluxetable}



\begin{thebibliography}{}

\bibitem[Bahcall et al. (1995)]{ba95} Bahcall, N. A., Lubin,
L. M. \& Dorman, V.  1995, \apjl, 447, L81

\bibitem[Bernstein et al. (2000)]{be00} Bernstein, G. M., Smith, D., Jarvis,
M. \& Fischer, P. 1999, in preparation.

\bibitem[Brainerd et al. (1996)]{br96} Brainerd, T. G., Blandford, R. D., \&
Smail, I. 1996, ApJ, 466, 623.

\bibitem[Budavari et al (1999)]{bu99} Budav\'ari, T., Szalay, A.S., Connolly,
A.J., Csabai, I. \& Dickinson, M.E., 1999, in preparation

\bibitem[Connolly et al. (1995)]{co95} Connolly, A.J., Csabai, I., Szalay, A.S.,
Koo, D.C., Kron, R.G. \& Munn, J.A., 1995, AJ, 110, 2655

\bibitem[Connolly et al. (1999)]{co99} Connolly, A.J., Budav\'ari, T., Szalay,
A.S. \& Csabai, I., 1999, to appear in "Photometric Redshifts and High Redshift
Galaxies", eds. R. Weymann, L. Storrie-Lombardi, M.  Sawicki \& R. Brunner,
(San Francisco: ASP Conference Series)

\bibitem[Csabai et al. (1999)]{cs99} Csabai, I., Connolly, A.J., Szalay, A.S.  \&
Budav\'ari, T., 1999, AJ, submitted 

\bibitem[Coleman et al. (1980)]{co80} Coleman, G.D., Wu., C.-C. \& Weedman, D.W.,
1980, ApJS, 43, 393. 

\bibitem[Dekel \& Lahav (1999)]{dl99} Dekel, A., \& Lahav, O. 1999, ApJ, 520,
24.

\bibitem[Dell'Antonio \& Tyson (1996)]{de96} Dell'Antonio, I.  P. \& Tyson, J. A.
1996, ApJ, 473, L17

\bibitem[Doi, M. et al. (2000)]{do00} Doi, M. et al. 2000, in preparation

\bibitem[Fan et al. (1999)]{fa99} Fan, Xiaohui, et al. 1999, AJ, 118, 1

\bibitem[Fukugita et al. (1996)]{fu96} Fukugita, M., Ichikawa, T., Gunn, J. E.,
Doi, M., Shimasaku, K., \& Schneider, D. P. 1996, AJ, 111, 1748.

\bibitem[Fukugita et al. (1999)]{fu99} Fukugita, M., Ichikawa, T. \& Sekiguchi,
M. 1999, in preparation.

\bibitem[Gould \& Villumsen (1994)]{go94} Gould, A. \& Villumsen, J. 1994,
\apjl, 428, L45.

\bibitem[Gilmore et al. (1990)]{gi90} Gilmore, G., King, I. R. \& van der
Kruit, P. 1990, ``The Milky Way as a Galaxy'', University Science Books, Mill
Valley California.

\bibitem[Griffiths, Casertano, Im, \& Ratnatunga (1996)]{gr96} Griffiths,
R. E., Casertano, S., Im, M. \& Ratnatunga, K. U. 1996, MNRAS, 282, 1159

\bibitem[Gunn et al (1998)]{gu98} Gunn, J., et al., 1998, AJ, 116, 3040.

\bibitem[Heyl et al. (1997)]{he97} Heyl, J., Colless, M., Ellis, R.S. \&
Broadhurst, T., 1997, MNRAS, 285, 613

\bibitem[Hudson, Gwyn, Dahle \& Kaiser (1998)]{hu98} Hudson, M.  J., Gwyn,
S. D. J., Dahle, H.  \& Kaiser, N.  1998, ApJ, 503, 531.

\bibitem[Kaiser (1992)]{ka92} Kaiser, N. 1992, ApJ, 388, 272.

\bibitem[Kaiser et al. (1995)]{ka95} Kaiser, N., Squires, G. \& Broadhurst,
T. 1995, ApJ, 449, 460.

\bibitem[Kassiola \& Kovner (1993)]{ka93} Kassiola, A, \& Kovner, I. 1993, ApJ,
417, 450.
 
\bibitem[Kent et al. (2000)]{ke00} Kent, S. M. et al. 2000, in preparation.

\bibitem[Koo, Gronwall, \& Bruzual (1993)]{kgb93} Koo, D.C., Gronwall, C.,
Bruzual, G.A. 1993, \apjl, 415, 272.

\bibitem[Luppino \& Kaiser (1997)]{lu97} Luppino, G. A. \& Kaiser, N. 1997,
ApJ, 475, 20

\bibitem[Lupton et al. (2000)]{lu00} Lupton, R. H. et al 2000, in preparation. 

\bibitem[McClelland \& Silk (1977)]{mc77} McClelland J., Silk J., 1977, ApJ,
217, 331

\bibitem[Mellier (1999)]{me99} Mellier, Y. 1999, ARAA, Vol. 37. 

\bibitem[Miralda-Escud\'{e} (1991)]{mi91} Miralda-Escud\'{e}, J. 1991, ApJ,
370, 1.

\bibitem[Miralda-Escud\'e (1996)]{mi96} Miralda-Escud\'e, J. 1996, in ``IAU
173: Astrophysical Applications of Gravitational Lensing'', eds. C. S. Kochanek
\& J. N. Hewitt, (Kluwer), p. 131.

\bibitem[Pen (1998)]{p98} Pen, U. 1998, ApJ, 504, 601.

\bibitem[Petravick et al. (2000)]{pe00} Petravick, D. et al. 2000, in
preparation.

\bibitem[Pier et al. (2000)]{pi00} Pier, J. R. et al. 2000, in preparation.

\bibitem[Sawicki et al (1997)]{sa97} Sawicki, M.J., Lin, H. \& Yee, H.K.C.,
1997, AJ, 113, 1

\bibitem[Schlegel et al. (1998)]{sc98} Schlegel, D. J., Finkbeiner, D. P., \&
Davis, M. 1998, ApJ, 500, 525.

\bibitem[Schneider \& Rix (1997)]{sc97} Schneider, P. \& Rix, H. 1997, ApJ,
474, 25.

\bibitem[Sheth \& Jain (1997)]{sh97} Sheth, R. \& Jain, B. 1997, MNRAS, 285, 231.

\bibitem[Siegmund et al (2000)]{si00} Siegmund et al. 2000, in preparation.

\bibitem[Stebbins, Mckay, \& Frieman (1996)]{st96} Stebbins, A., Mckay, T., \&
Frieman, J. 1996, in ``IAU 173: Astrophysical Applications of Gravitational
Lensing'', eds. C. S. Kochanek \& J. N. Hewitt, (Kluwer), p. 75.

\bibitem[Tucker et al. (2000)]{tu00} Tucker, D. L. et al. 2000, in preparation.

\bibitem[Tyson et al. (1984)]{ty84} Tyson, J. A., Valdes, F., Jarvis, J. F.,
Mills, A. P., JR. 1984, ApJ, 281, L59.

\bibitem[Tyson, Kochanski \& Dell'Antonio (1998)]{ty98} Tyson, J. A. ,
Kochanski, G. P. \& Dell'Antonio, I. P. 1998, \apjl, 498, L107

\bibitem[Uomoto et al. (2000)]{uo00} Uomoto, A. et al. 2000, in preparation

\bibitem[van Waerbeke (1998)]{vW98} van Waerbeke, L. 1998, A\&A, 334, 1.

\bibitem[Williams et al. (1996)]{wi96} Williams et al. 1996, AJ, 112, 1335.

\bibitem[Zaritsky \& White (1994)]{za94} Zaritsky, D. \& White, S. D. M. 1994,
ApJ, 435, 599. 

\end{thebibliography}
\end{document}